\documentclass[12pt, draftclsnofoot, onecolumn]{IEEEtran}
\usepackage{cite}
\usepackage{algorithm2e}
\usepackage{amssymb}
\usepackage{enumerate}
\usepackage{graphicx}
\usepackage{amssymb}
\usepackage{amsbsy}
\usepackage{float}
\usepackage{balance}
\usepackage{url}
\usepackage{array}
\usepackage{varioref}
\usepackage[normalem]{ulem}

\usepackage{color}
\usepackage{empheq}
\usepackage{flushend}
\begin{document}

\title{\textcolor{black}{Multicast Multigroup Precoding and User Scheduling for Frame-Based Satellite Communications} }
\author{Dimitrios Christopoulos, \IEEEmembership{Member, IEEE}, Symeon Chatzinotas, \IEEEmembership{Senior Member, IEEE,} and \\ Bj\"orn Ottersten, \IEEEmembership{Fellow, IEEE}
\thanks{The authors are with the SnT, University of Luxembourg. Email: \textbraceleft dimitrios.christopoulos, symeon.chatzinotas, bjorn.ottersten\textbraceright@uni.lu. This work was   partially supported by the National Research Fund, Luxembourg under the projects  ``$\mathrm{ CO^{2}SAT}$' and ``$\mathrm{ SeMIGod:}$''. Part of this work has been presented at the IEEE GlobeCom 2014 conference.}}
\markboth{IEEE transactions in Wireless Communications (\MakeLowercase{preprint} \copyright IEEE)}{Christopoulos \MakeLowercase{\text
it{et al.}}: Multicast Multigroup Precoding and User Scheduling for Frame-Based Satellite Communications}
\maketitle
%
\begin{abstract}
The present work focuses on the forward link of a broadband multibeam  satellite system that aggressively reuses the user link frequency resources. Two fundamental practical challenges, namely the need to frame multiple users per transmission and the per-antenna transmit power limitations, are addressed. To this end, the so-called frame-based precoding problem is optimally solved using the principles of physical layer multicasting to multiple co-channel groups under per-antenna constraints. In this context, a novel optimization problem that aims at maximizing the system sum rate under individual power constraints is proposed. Added to that, the formulation is further extended to include availability constraints. As a result, the high gains of the sum rate optimal design are traded off to satisfy the stringent availability requirements of satellite systems. Moreover, the throughput maximization with a granular spectral efficiency versus $\mathrm{SINR}$ function, is formulated and solved. Finally,  a multicast-aware user scheduling policy, based on the channel state information, is developed. Thus, \textcolor{black}{substantial multiuser diversity gains }  are gleaned. Numerical results over a realistic simulation environment exhibit as much as 30\% gains over conventional systems, even for 7 users per frame, without modifying the framing structure of legacy communication standards.
 \end{abstract}

\section{Introduction \& Related Work}

Aggressive frequency reuse schemes have shown to be the most promising way towards spectrally efficient,  high-throughput wireless communications. In this context, linear precoding, a transmit signal processing technique that  exploits the offered spatial degrees of freedom of a multi-antenna transmitter, is brought into play to manage interferences. Such interference mitigation techniques and subsequently full frequency reuse configurations, are enabled by the availability  of channel state information ($\mathrm{CSI}$) at the transmitter.

In fixed broadband multibeam    satellite communications (satcoms),  the relatively slow channel variations facilitate the channel acquisition process. Therefore,  such scenarios emerge as the most promising use cases of  full frequency reuse configurations. Nevertheless, the incorporation of linear precoding techniques is inhibited by the inherent characteristics of the satellite system \cite{Christopoulos2013AIAA,Taricco2014}. The present contribution focuses on two fundamental constraints stemming from the practical system implementation. Firstly,   the  framing structure of  satcom standards, such as  the second generation digital video broadcasting for satellite standard $\mathrm{DVB-S2}$\cite{DVB_S2_standard} and its most recent extensions $\mathrm{DVB-S2X}~$\cite{DVB_S2X_standard}, inhibit scheduling a single user per transmission.  Secondly,  non-flexible on-board payloads prevent power sharing between  beams.

\begin{figure*}[!htp]
\centering
\includegraphics[width = 0.85\textwidth]{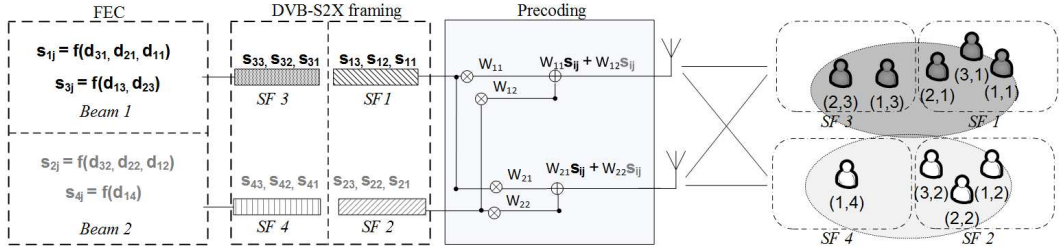}\\
\caption{\textcolor{black}{Frame-based precoding in $\mathrm{DVB-S2X}$. \textcolor{black}{Function $f(\cdot)$ denotes the $\mathrm{FEC}$ coding operation over the data $d_{xy}$  that are uniquely addressed   to   user ${x}$ of beam ${y  }$, as identified in the right side of the plot.   Consequently, the $j$-th transmitted  symbol   ${s_{ij}}$,  belonging to the  ${{i}}$-th      superframe ($\mathrm{SF}$), contains an encoded bit-stream that  needs to be received by all co-scheduled users. In $\mathrm{SF}$s 3 and 4,  different number of users are co-scheduled.}}  }
\label{fig: proposed system model}
\end{figure*}

Focusing on the first practical constraint, the physical layer design of $\mathrm{DVB-S2}$ \cite{DVB_S2_standard} has been optimized  to cope with the 
  noise limited, with excessive propagation delays and intense fading phenomena, satellite channel. Therefore, long forward error correction ($\mathrm{FEC}$) codes    and   fade mitigation techniques that rely on an adaptive link layer design (adaptive coding and modulation -- $\mathrm{ACM}$) have been employed. The latest evolution of $\mathrm{DVB-S2X}$, \textcolor{black}{through its  --synchronous over the multiple beams-- superframes (cf. annex E of \cite{DVB_S2X_standard})}, allows for the incorporation of the aforementioned interference mitigation techniques (cf. annex C of \cite{DVB_S2X}). A small-scale example of the application of linear precoding methods within the $\mathrm{DVB-S2X}$ standard is depicted in  Fig. \ref{fig: proposed system model}. Clearly,  the underlying framing structure   hinders the calculation of a precoding matrix on a user-by-user basis. During one transmission period, one frame per beam accommodates
a different  number of users,  each with different  data requirements. Added to that, the application of $\mathrm{FEC}$ block coding over the entire frame requires that co-scheduled users   decode the entire frame and then extract \textcolor{black}{the data they need}. Also, the unequal data payloads amongst users simultaneously served in different beams further complicates the joint processing of the multiple streams. Consequently, despite the capacity achieving channel based precoding \cite{Christopoulos2012}, practical system implementations emanate  the consideration of precoding  on a frame-by-frame basis.
\textcolor{black}{The notion of frame-based precoding is presented in more detail in \cite{Christopoulos2013AIAA,Taricco2014}. }



From a signal processing perspective, physical layer ($\mathrm{PHY}$) multicasting to multiple co-channel groups \cite{Karipidis2008} can provide the theoretically optimal precoders when a multi-antenna transmitter conveys independent sets of common data to distinct
groups of users. This scenario is known as $\mathrm{PHY}$  multigroup multicast beamforming (or equivalently precoding). The optimality of the  multicast multigroup precoders for frame-based precoding is intuitively clear, under the following considerations. In
multicasting, the same symbol is transmitted to multiple receivers.  This is the fundamental assumption of frame-based precoding as well, since the symbols of one frame, regardless of the information they convey, are addressed to multiple users. These users need to receive the entire frame, decode it \textcolor{black}{and then extract  information that is relevant to them}.
\textcolor{black}{ The connection between  $\mathrm{PHY}$  multigroup multicast beamforming (precoding) and frame-based precoding  was firstly established in \cite{Christopoulos2014ASMS}.}


 The second practical  constraint tackled in the present work includes   a maximum limit on the per-antenna transmitted power.    Individual per-antenna amplifiers prevent power sharing amongst the antennas of the future full frequency reuse compatible satellites. On board flexible amplifiers, such as multi-port amplifiers and flexible  traveling wave tube amplifiers \cite{Zheng2012}, come at high costs. Also, power sharing is impossible in distributed antenna systems ($\mathrm{DAS}$), such as constellations of cooperative satellite systems (e.g. dual satellite systems \cite{Christopoulos2013a} or swarms of nano-satellites).

  Enabled by the incorporation of linear precoding in DVB-S2X,  an example of a full frequency reuse transmission chain is depicted in Fig. 2. The optimal, in a throughput maximizing sense, precoding matrix, combined with a low complexity user scheduling algorithm
will be presented in the remaining parts of this work.

\begin{figure}
 \centering
 \includegraphics[width=0.72\columnwidth]{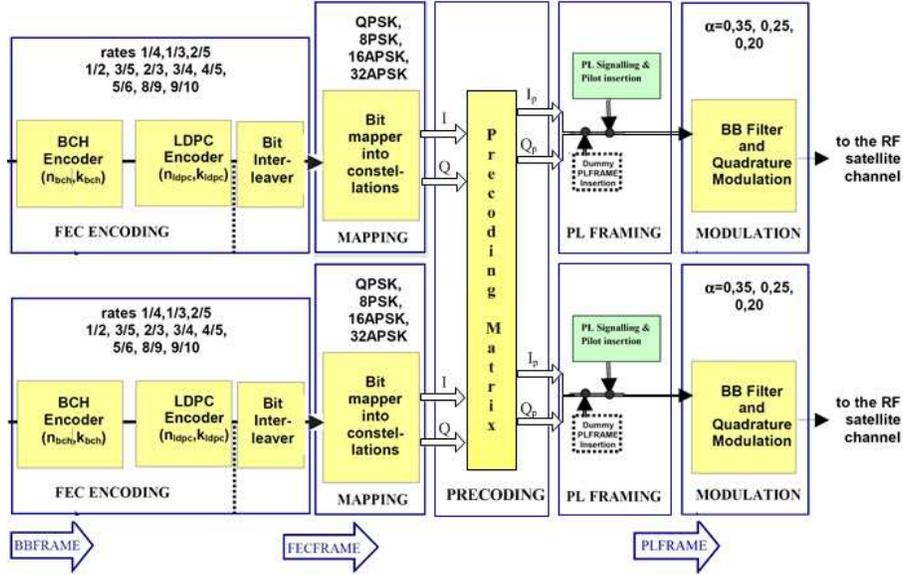}\\
  \caption{Transmitter functional block diagram, based on   DVB-S2 \cite{DVB_S2_standard}, extended to incorporate advanced interference mitigation techniques.}
\label{fig: precoding}
\end{figure}
\subsection{Related Work}
In the  $\mathrm{PHY}$ multigroup multicast precoding literature, two fundamental optimization criteria, namely the sum power minimization under specific Quality of Service ($\mathrm{QoS}$) constraints and the maximization of the minimum $\mathrm{SINR}$ ($\max \min$ fair criterion) have been considered in \cite{Sidiropoulos2006,Karipidis2008,Silva2009} under a $\mathrm{SPC}$. Extending these works, a consolidated solution for the weighted $\max \min$ fair multigroup multicast beamforming under $\mathrm{PAC}$s has been derived in \cite{Christopoulos2014,Christopoulos2014ICC}. To this end, the well established tools of Semi-Definite Relaxation ($\mathrm{SDR}$) and Gaussian randomization were combined with bisection to obtain highly accurate and efficient solutions.

The fundamental attribute of  multicasting, that is    a  single transmission to be addressed to a group of users, constrains the system performance according to the worst user.  Therefore, the maximization of the minimum $\mathrm{SINR}$ is the most relevant problem and the fairness criterion is imperative\cite{Christopoulos2014}. When advancing to multigroup multicast systems, however, the service levels between different groups can be adjusted towards achieving some other optimization goal. The sum rate maximization  ($\max \mathrm{SR}$) problem in the multigroup multicast context was initially considered in  \cite{Kaliszan2012} under $\mathrm{SPC}$. Therein, a heuristic iterative algorithm  based on the principle of decoupling the beamforming design and the power allocation problem was proposed. In more detail, the $\mathrm{SPC}$ max sum rate problem was solved using a two step optimization algorithm. The first step was based on   the  $\mathrm{QoS}$ multicast beamforming problem of \cite{Karipidis2008}, as iteratively solved with input $\mathrm{QoS}$ targets defined by the worst user per group in the previous iteration. The derived precoders push all the users of the group closer to the worst user thus saving power. The second step of the algorithm consisted \textcolor{black}{of the gradient} based power reallocation methods of \cite{stanczak2009fundamentals}.  Hence, a power redistribution  takes place via the sub-gradient method \cite{stanczak2009fundamentals} to the end of  maximizing the system sum rate.

In a realistic system design, besides the optimal transmission method, the  need to schedule a large number users over subsequent in time transmissions is of substantial importance.  In the context of multiuser multiple input multiple output ($\mathrm{MU-MIMO}$) communications, user scheduling has shown great potential in maximizing the system throughput performance. In \cite{dimic2005,Yoo2006b}, low complexity user scheduling algorithms  allowed for the channel capacity approaching performance of linear precoding methods when the number of available  users grows large. The enabler for these algorithms is the exact knowledge of the $\mathrm{CSI}$. Motivated by these results and by acknowledging that  the large number  of users served by one satellite can offer significant multiuser diversity gains, channel based  user scheduling over satellite is herein proposed. Further supporting this claim, \textcolor{black}{  the diverse multiuser satellite environment}  was   exploited towards approaching the information theoretic channel capacity bounds in \cite{Christopoulos2013a}. Therein, user scheduling  methods where extended to account for adjacent transmitters and applied in a multibeam satellite scenario, exhibiting the importance of scheduling for satcoms.
In the present work, drawing intuitions from the frame-based design,  multicast-aware user scheduling algorithms are derived. These algorithms, as it will be shown, exploit the readily available $\mathrm{CSI}$, to glean the \textcolor{black}{multiuser diversity gains} of   satellite systems.

\textcolor{black}{Different from the aforementioned works, the   sum rate maximization under   $\mathrm{PAC}$s has only been considered in \cite{Christopoulos2014Globecom}. Herein, this principle is used as a stepping stone for the incremental development of elaborate optimization algorithms that solve problems inspired by the needs of   frame-based precoding over satellite.} The contributions  are summarized in the following points:
 \begin{itemize}
\item
The $\max$ $\mathrm{SR}$ multigroup multicast problem under $\mathrm{PAC}$s is formulated and solved.
\item
 The above $\max$ $\mathrm{SR}$ problem is extended to account  for  minimum rate constraints ($\mathrm{MRC}$s).
\item A\ novel modulation aware $\max$ $\mathrm{SR}$ optimization that considers the  discretized throughput function of the receive useful signal power is proposed  and heuristically solved.
\item  A low complexity,  $\mathrm{CSI}$ based, user scheduling algorithm that considers the multigroup multicast nature of the frame-based precoding  system is envisaged. \item The  developed techniques are evaluated over a  multibeam, full frequency reuse satellite  scenario. \end{itemize}
The rest of the paper is structured as follows.   Section \ref{sec: System Model} models the multigroup multicast system. Based on this model, the $\max \mathrm{SR}$, multigroup multicast optimization problem is formulated and solved in Sec. \ref{sec: SR}. Extending this optimization, system dependent problems are tackled in Sec.  \ref{sec: SRM}. Further on, user scheduling is discussed in Sec. \ref{sec: scheduling}. Finally, in Sec. \ref{sec: performance}, the performance of the derived algorithms is evaluated, while   Sec. \ref{sec: conclusions} concludes the paper.


{\textit{Notation}: In the remainder of this paper, bold face lower case and upper case characters denote column vectors  and matrices, respectively. The operators \(\left(\cdot\right)^\text{T}\), \(\left(\cdot\right)^\dag\), $|\cdot|$, $\mathrm{Tr}\left(\cdot\right)$ and \(||\cdot||_2, \) correspond to   the transpose, the conjugate transpose,  the absolute value, the trace and the \textcolor{black}{Euclidean norm} operations,  while $[\cdot]_{ij}  $  denotes the $i, j$-th element of a matrix. An $x$-element  column vector of ones  is denoted as $\mathbf 1_{x}$. \textcolor{black}{Finally},  ${\emptyset}$ denotes an empty set}.

\section{System Model}\label{sec: System Model}
 The focus is on a single broadband multibeam satellite transmitting to multiple single antenna users.  Let   $N_t$ denote the number of transmitting elements, \textcolor{black}{which for the purposes of the present work, are considered} equal to the number of beams (one feed per beam assumption)  and  $N_{u}$ the  total number of users simultaneously served.  The received signal at the $i$-th user will read as $y_{i}= \mathbf h^{\dag}_{i}\mathbf x+n_{i},$
where \(\mathbf h^{\dag}_{i}\) is a \(1 \times N_{t}\) vector composed of the channel coefficients (i.e. channel gains and phases) between the \(i\)-th user and the  \(N_{t}\) antennas of the transmitter, \(\mathbf x\) is the \(N_{t} \times 1\)  vector of the transmitted symbols and  \(n_{i}\) is the complex circular symmetric (c.c.s.) independent identically distributed (i.i.d) zero mean  Additive White Gaussian Noise ($\mathrm{AWGN}$),  measured at the \(i\)-th user's receiver.
 \textcolor{black}{Herein, for simplicity, the noise will be normalized to one and the impact of noise at the receiver side will be incorporated in the channel coefficients, as will be shown in the following (Sec. II.A eq. (4) ). }

Let us assume that a total of $ N_{t}$ multicast groups are realized where  $\mathcal{I} = \{\mathcal{G}_1, \mathcal{G}_2, \dots  \mathcal{G}_{N_{t}}\}$ the collection of   index sets and $\mathcal{G}_k$ the set of users that belong to the $k$-th multicast group, $k \in \{1\dots {N_{t}} \}$. Each user belongs to only one frame (i.e. group), thus $\mathcal{G}_i\cap\mathcal{G}_j=$\O ,$  \forall i,j \in \{1\cdots {N_{t}}\}$, \textcolor{black}{while $\rho = N_u/N_t$ denotes the number of users per group.} Let $\mathbf w_k \in \mathbb{C}^{N_t \times 1}$ denote the precoding weight vector applied to the transmit antennas to beamform towards the $k$-th group of users.
By collecting all user channels in one channel matrix, the general linear signal model in vector form reads as $ \mathbf y = \mathbf {H}\mathbf x + \mathbf n = \mathbf {H}\mathbf {Ws} + \mathbf{ n}$, where $ \mathbf {y \text{ and }  n } \in \mathcal{\mathbb{C}}^{N_{u}}$, $\mathbf {x} \in \mathbb{C}^{N_{t}}$ and $ \mathbf {H} \in \mathbb{C}^{N_{u} \times N_t}$.  Since, the frame-based precoding imposes  a single precoding vector for multiple users, the matrix  will include as many precoding vectors (i.e columns) as the number of multicast groups. This is the number of transmit antennas, since  one frame per-antenna is assumed. Also,  the symbol vector includes a single equivalent symbol for each frame i.e. $\mathbf {s} \in \mathbb{C}^{{N_{t}}}$, inline with the multicast assumptions. Consequently, a   square precoding matrix is realized, i.e.
$ \mathbf {W} \in \mathbb{C}^{N_t  \times {N_{t}}}$. The assumption of independent information transmitted to different frames implies that the symbol streams $\{s_k\}_{k=1}^{N_{t}}$ are mutually uncorrelated. \textcolor{black}{Also, the average power of the transmitted symbols is assumed normalized to one.} Therefore, the total power radiated from the antenna array is equal to
\begin{align}
P_{tot} = \sum_{k=1}^{N_{t}} \mathbf w_k^\dag \mathbf w_k= \mathrm{Trace\left( \mathbf {WW}^\dag\right)},\label{eq: SPC}
\end{align}
where $\mathbf {W}= [\mathbf w_1, \mathbf w_2, \dots\mathbf w_{N_{t}}].$
The power radiated by each
antenna element is  a  linear combination of all precoders and reads as
\cite{Yu2007}\begin{align}\label{eq: PAC}
P_n = \left[\sum_{k=1}^{N_{t}} \mathbf w_k \mathbf w_k^\dag \right]_{nn} =\left[ \mathbf {WW}^\dag\right]_{nn},
\end{align}
where $n \in \{1\dots  N_t\}$ is the antenna index.
The fundamental difference between the $\mathrm{SPC}$  of \cite{Karipidis2008} and the proposed $\mathrm{PAC}$ is clear in  \eqref{eq: PAC}, where instead of one,  $N_t$ constraints are realized, each one involving all the precoding vectors.

\subsection{Multibeam Satellite Channel }\label{sec: channel}
\begin{figure}[!htp]
 \centering
 \includegraphics[width=0.65\columnwidth]{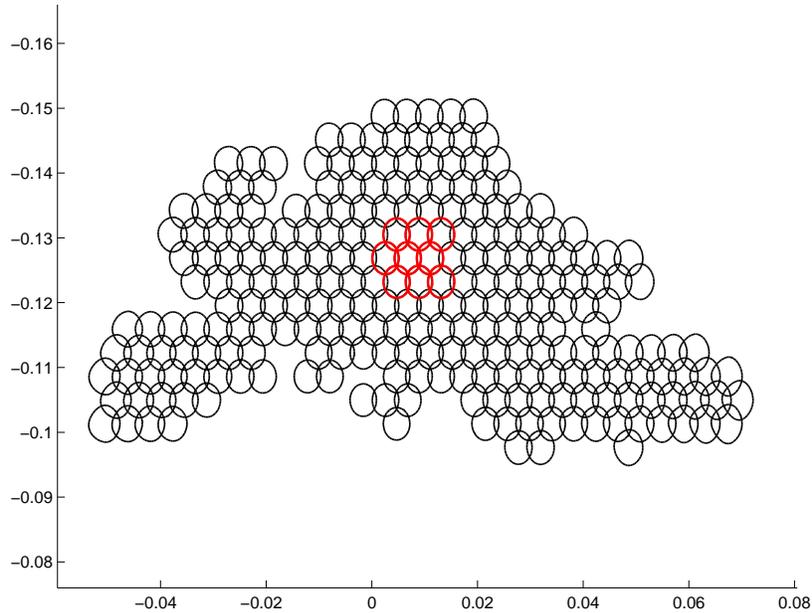}\\
  \caption{Beam pattern covering Europe, provided by \cite{satnex}, with the nine beams considered herein highlighted. }
\label{fig: CA}
\end{figure}

The above general system model is  applied over a multibeam satellite channel explicitly defined as follows. A 245 beam pattern that covers Europe is employed\cite{satnex}. For the purposes of the present work, only a subset of the 245 beams will be considered, as presented in Fig. \ref{fig: CA}. Such a consideration is in line with the \textcolor{black}{multiple gate-way (multi-GW) assumptions}  of large multibeam systems\cite{Zheng2011c}. However, the effects of interference from adjacent clusters is left for future investigations.    A complex channel matrix that models the link budget of each user as well as the phase rotations induced by the signal propagation  is employedin the standards of \cite{satnex}, \cite{Zheng2012} and \cite{Christopoulos2014ASMS}.
  In more detail, the total channel matrix    \( \mathbf{H}\in \mathbb{C}^{N_u \times N_t }\)
 is generated  as
\begin{equation}\label{eq: phase assumptions}
\mathbf{ H}=\mathbf{\Phi}\mathbf{B},
\end{equation}
and includes the multibeam antenna pattern (matrix $\mathbf{B}$) and the signal phase due to different propagation paths between the users (matrix $\mathbf{\Phi}$). The real matrix $\mathbf{B} \in \mathbb{R}^{N_u\times N_t}$ models the satellite antenna radiation pattern, the path loss, the receive antenna gain and  the noise power.  Its $i,j$-th entry is given by \cite{satnex}:
\begin{equation}\label{eq: beam_gain}
b_{ij}=\left(\frac{\sqrt{G_RG_{ij}}}{4\pi(d_k\cdot\lambda^{-1})\sqrt{\kappa T_{cs}B_u}}\right),
\end{equation}
with \(d_k\) the distance between the \textcolor{black}{$i$-th} user and the satellite (slant-range), \(\lambda\) the wavelength, \(\kappa\) the Boltzman constant,  \(T_{cs}\) the clear sky noise temperature of the receiver, $B_u$ the user link bandwidth, $G_R$ the receiver antenna gain and $G_{ij}$ the multibeam antenna gain between the $i$-th single antenna user and the $j$-th on board antenna (= feed).
Hence, the beam gain for each satellite antenna-user pair, depends on the antenna pattern and on  the user position.

An inherent characteristic of the multibeam satellite channel is the high correlation of signals at the satellite side. Thus a common assumption in multibeam channel models is that each user will have the same phase between all transmit antennas due to the long propagation path \cite{Zheng2012}. The identical phase assumption between one user and all transmit feeds is supported by the relatively small distances between the transmit antennas and the long propagation distance of all signals to a specific receiver.
 Hence, in \eqref{eq: phase assumptions} the diagonal square matrix
$\mathbf \Phi$
is generated as
$[\mathbf \Phi]_{xx} = e^{\mathrm j \phi_x}, \forall \ x = 1\dots N_{u}$
where
$\phi_{x}$
is  a uniform random variable in
$[2 \pi, 0) $
and
$[\mathbf \Phi_{} ] _{xy}=0, \ \forall \ x \neq y $.
\subsection{Average User Throughput}
Based on the above link budget considerations, the achievable average user throughput is normalized   over the  number of beams, in order to provide a  metric comparable with multibeam systems of any size.     Therefore, the average user throughput, $R_{avg}$ as will be hereafter referred to, is given as
\begin{align}\label{eq: throughput}
R_{avg} =\frac{2B_u}{1+\alpha} \frac{1}{{N_{t}}}\sum_{k=1 }^{N_{t}}f_{\mathrm{DVB-S2X}}\left(\min_{i \in\mathcal{G}_k}\left\{\mathrm{SINR}_i\right\}, \mathbf t \right  ),
\end{align}
in [Gbps/beam], where all parameters are defined in Tab. \ref{tab: simulation params} of Sec. VI.
In \eqref{eq: throughput},  the spectral efficiency function $f_{\mathrm{DVB-S2X}}$ receives as input each users $\mathrm{SINR}$ as well as a threshold vector $\mathbf t$.  Then, $f_{\mathrm{DVB-S2X}}$ performs a  rounding of the input  $\mathrm{SINR}$  to the closest lower floor given by the threshold vector $\mathbf t$ and outputs the corresponding spectral efficiency in [bps/Hz]. This operation is denoted as $\left\lfloor \cdot\right\rfloor_{\mathbf t}$. The mapping of receive $\mathrm{SINR}$ regions to a spectral efficiency achieved by a respective modulation and coding ($\mathrm{MODCOD}$) scheme is explicitly defined in the latest evolution of the satcom standards \cite{DVB_S2X_standard}. It should also be noted, that  the conventional four color frequency reuse calculations are based on the exact same formula, with the only modifications  being the  input  $\mathrm{SINR}$, calculated under conventional four color  reuse  pattern and with  the pre-log factor reduced by four times, equal to the conventional fractional frequency reuse \cite{satnex}.

 \section{Sum Rate Maximization }\label{sec: SR}

For the precoding design, optimal multigroup multicast precoders under per-antenna constraints are proposed to maximize the throughput of the multibeam satellite system.
 \textcolor{black}{The design of throughput maximizing optimal precoders  is a complicated problem without an explicit solution even for the unicasting case \cite{Bjornson2014}. When advancing to multicasting assumptions, the structure of the problem becomes even more involved, as already explained \cite{Sidiropoulos2006}.
Consequently, the present work builds upon the heuristic methods of \cite{Kaliszan2012,stanczak2009fundamentals}. }

Since a  multigroup multicasting scenario entails the flexibility  to maximize the total  system rate by providing different service levels amongst groups, the multigroup multicast $\max \mathrm{SR}$ optimization  aims  at increasing the minimum $\mathrm{SINR}$ within each group while in parallel maximizing the sum of the rates of all groups.
Intuitively, this can be accomplished by reducing the $\mathrm{SINR}$ of  users with better conditions  than the worst user of their group. Also, groups that contain compromised users might need to be turned of, hence driving their users  to service unavailability, in order to save power resources and degrees of freedom. As a result, power is not consumed for the mitigation of poor channel conditions. Any remaining power budget is then reallocated to well conditioned and balanced in terms of performance groups.
\subsection{Per-antenna Power Constrained Optimization }
This section focuses on the per-antenna power constrained $\max \mathrm{SR}$ problem,  formally defined as
  \begin{empheq}[box=\fbox]{align}
\mathcal{SR:}\    \max_{\   \ \{\mathbf w_k \}_{k=1}^{N_t}}  &\sum_{i=1}^{N_u} \log_2\left(1+\gamma_i \right) & \notag\\
\mbox{subject to: } & \gamma_i  = \min_{m\in \mathcal G_k}\frac{|\mathbf w_k^\dag \mathbf h_m|^2}{\sum_{l\neq k}^{N_t} |\mathbf w_l^\dag\mathbf h_m|^2+\sigma_m^2 }, &\label{const: SR SINR}\\
&\forall i \in\mathcal{G}_k, k,l\in\{1\dots {N_t}\},\notag\\
 \text{and to: } & \left[ \sum^{N_{t}}_{k=1} \mathbf w_k\mathbf w_k^\dag  \right]_{nn}  \leq P_n, \label{eq: SR PAC}\\%
 &\forall n\in \{1\dots N_{t}\}.\notag
 \end{empheq}
Problem $\mathcal{SR}$ receives as input the channel matrices as well as the per-antenna power constraint vector $\mathbf p_{ant} = [P_1, P_2\dots P_{N_t}]$. \textcolor{black}{Following the notation of \cite{Karipidis2008} for ease of reference}, the  optimal objective value of $ \mathcal{SR}$ will be denoted as $c^*=\mathcal{SR}(  \mathbf p_{ant})$ and the associated optimal point as $\{\mathbf w_k^\mathcal{SR}\ \}_{k=1}^{{N_t}}$. The novelty of the $\mathcal{SR}$  lies in the $\mathrm{PAC}s$, i.e.
\eqref{eq: SR PAC} instead of the conventional $\mathrm{SPC}$
 proposed in \cite{Kaliszan2012}. Therein, to solve the elaborate $\max \mathrm{SR}$ under a $\mathrm{SPC}$ problem, the decoupling of the precoder calculation and the power loading over these vectors was considered. The first problem was solved based on the solutions of \cite{Karipidis2008} while the latter on sub-gradient optimization methods \cite{stanczak2009fundamentals}.
 To the end of solving the novel $\mathcal{SR}$ problem, a  heuristic algorithm is proposed herein. Different than in \cite{Kaliszan2012},   the new algorithm calculates the per-antenna power constrained precoders by utilizing recent results \cite{Christopoulos2014}. Also, modified  sub-gradient optimization methods are proposed to take into account the $\mathrm{PAC}$s. More specifically, instead of solving the $\mathrm{QoS}$ sum power minimization problem of \cite{Karipidis2008}, the proposed algorithm  calculates the $\mathrm{PAC}$ precoding vectors by solving the following  problem \cite{Christopoulos2014} that reads as\begin{empheq}[box=\fbox]{align}
\mathcal{Q:} \min_{\ r, \ \{\mathbf w_k \}_{k=1}^{N_t}}  &r& \notag\\
\mbox{subject to: } & \frac{|\mathbf w_k^\dag \mathbf h_i|^2}{\sum_{l\neq k }^{N_t} | \mathbf w_l^\dag\mathbf h_i|^2 + \sigma^2_i}\geq \gamma_i, \label{const: Q SINR}\\
&\forall i \in\mathcal{G}_k, k,l\in\{1\dots{N_t}\},\notag\\
\text{and to:} &\frac{1}{P_n} \left[\sum_{k=1}^{N_t}  \mathbf w_k\mathbf w_k^\dag \right]_{nn} \leq  r,\\
& \forall n\in \{1\dots N_{t}\}, \notag
 \end{empheq}
 where $r\in\mathbb{R}^+$. Problem $\mathcal{Q }$ receives as input  the  $\mathrm{SINR}$ target  vector $\mathbf g = [\gamma_1,\gamma_2, \dots \gamma_{N_u}]$, that is the individual  $\mathrm{QoS}$ constraints of each user,   as well as the per-antenna power constraint vector $\mathbf p_{ant} .$ Let the  optimal objective value of $ \mathcal{Q}$  be denoted as $r^*=\mathcal{Q}(\mathbf g,  \mathbf p_{ant})$ and the associated optimal point as $\{\mathbf w_k^\mathcal{Q}\ \}_{k=1}^{{N_t}}$. This problem is solved using the well established methods of $\mathrm{SDR}$ and Gaussian randomization \cite{Luo2010}. A more detailed description  of the solution of $ \mathcal{Q}$ can be found in \cite{Christopoulos2014ICC,Christopoulos2014} and is herein omitted for conciseness.

To proceed with the power reallocation step, let us rewrite the precoding vectors calculated from $\mathcal{Q }$  as $\{\mathbf w_k^\mathcal{Q} \}_{k=1}^{{N_t}}=\{\sqrt{p_k}\mathbf  v_k \}_{k=1}^{{N_t}}$ with \textcolor{black}{$ ||\mathbf v_k ||^2_2=1 $} and $\mathbf p =\left[p_1\dots p_k\right] $. By this normalization, the beamforming problem can be decoupled into two problems. The calculation of the beamforming directions, i.e. the normalized $\{\mathbf  v_k \}_{k=1}^{{N_t}}$, and the power allocation over the existing groups, i.e. the calculation of $\mathbf p_k$. Since the exact solution of $\mathcal{SR}$ is not straightforwardly obtained, this decoupling allows for a two step optimization. Under general unicasting assumptions, the $\mathrm{SR}$ maximizing power allocation with fixed beamforming directions is a convex optimization problem \cite{stanczak2009fundamentals}. Nonetheless, when multigroup multicasting is considered, the cost function   $C_{\mathcal{SR}} = \sum_{k=1 }^{N_t} \log\left(1+\min_{i \in\mathcal{G}_k}\left\{\mathrm{SINR}_i\right\}\right  ) . $ is no longer differentiable due to the $\min_{i \in\mathcal{G}_k}$ operation and one has to adhere to sub-gradient solutions\cite{Kaliszan2012}.
\textcolor{black}{\textcolor{black}{What is more, as in detail explained in \cite{Kaliszan2012}, the cost function needs to be  continuously differentiable, strictly increasing, with a log-convex inverse function. Nevertheless,  this is not the case for $\mathcal{SR}$. }  \textcolor{black}{Towards providing a heuristic solution to an involved problem without known optimal solution,  an optimization over the logarithmic power vector  $\mathbf s =\{ s_{k}\}_{k=1}^{N_t}= \{\log  p_{k}\}_{k=1}^{N_t} $, will be considered  in the standards of \cite{Kaliszan2012}. Therein, the authors\ employ a function $\phi$ that satisfies the above assumptions to approximate the utility function of $\mathcal{SR}$. For more information on function $\phi$ and the suggested approximation, the reader is directed to\cite{Kaliszan2012}. It should be noted that the heuristic nature of this solution does not necessarily guarantee convergence to a global optimum. Albeit this, and despite being sub-optimal in the max sum rate sense, the heuristic solutions attain a good performance, as shown in \cite{Kaliszan2012,stanczak2009fundamentals} and in the following.}} Consequently, in the present contribution,  the power loading is achieved via the sub-gradient method\cite{stanczak2009fundamentals}, under specific modifications over \cite{Kaliszan2012} that are hereafter described.

The proposed algorithm, presented  in Alg. \ref{Alg: MSR PAC}, is an iterative two step  procedure. In each step, the $\mathrm{QoS}$ targets $\mathbf g$ are calculated as the minimum target per group of the previous iteration, i.e. $\gamma_i = \min_{i \in\mathcal{G}_k}\left\{\mathrm{SINR}_i\right\}, \forall i \in\mathcal{G}_k, k\in\{1\dots {N_t}\} $. Therefore,  the new precoders require equal or less power to achieve the  same system sum rate. Any remaining power is then redistributed amongst the groups to the end of maximizing the total system throughput, via the sub-gradient method \cite{stanczak2009fundamentals}.
Focusing of the later method and using the logarithmic power vector  $\mathbf s =\{ s_{k}\}_{k=1}^{N_t}= \{\log  p_{k}\}_{k=1}^{N_t} $, the sub-gradient search method is given as
  \begin{align}
 \mathbf s{(t+1)}=\prod_{\mathbb P}\left[\mathbf s{(t)} - \delta(t) \cdot \mathbf{ r}(t)\right], \label{eq: subgrad}
   \end{align}
   where $\prod_{\mathbb P}[\mathbf x]$ denotes the projection operation of point $\mathbf x \in \mathbb{R}_{N_t}$ onto the set $\mathbb P\subset \mathbb  R^+_{N_t}$. The parameters $\delta(t) $  and $\mathbf{ r}(t)$ are the step of the search and the sub-gradient of the  $\mathcal{SR}$ cost function at the point $\mathbf s(t)$, respectively. The number of iterations this method runs, denoted as $t_{max}$, is predefined. The projection operation,  i.e. $\prod_{\mathbb P}[\cdot]$, constrains each iteration of the sub-gradient to the feasibility set of the $\mathcal{SR}$ problem. The analytic calculation of $\mathbf r(t) $ follows the exact steps of \cite{Kaliszan2012,stanczak2009fundamentals} and is herein omitted for shortness.
In order to account for the more complicated  $\mathrm{PAC}$s   the projection over a per-antenna power constrained set is  considered as follows. The set of $\mathrm{PAC}$s  can be defined as
\begin{align}
\mathbb P = \left \{\mathbf p _{}\in\mathbb{R}^+_{N_t}| \left[\textcolor{black}{\sum_{k=1}^{N_t}   p_k \mathbf v_k  \mathbf v_k^\dag}  \right]_{nn}  \leq P_n  \right\},\label{eq: proj}
\end{align}
where the elements of the power vector $\mathbf p_{}=\exp(\mathbf s)$ represent the power allocated to each group. It should be stressed that this power is inherently different from the power transmitted by each antenna $\mathbf{p}_{ant}\in\mathbb{R}^+_{N_t}$. The connection between $\mathbf{p}_{ant}$ and $\mathbf{p}$ is given by the normalized beamforming vectors as easily observed in   \eqref{eq: proj}. Different from the sum power constrained solutions of  \cite{Kaliszan2012}, the per-antenna constrained projection problem is given by
\begin{empheq}[box=\fbox]{align}
\mathcal{P:} \min_{\mathbf p}  &||\mathbf p -\mathbf x||^2_2& \notag\\
\mbox{subject to :} &  \left[ \textcolor{black}{\sum^{N_t}_{k=1} p_k \mathbf v_k \mathbf v_k^\dag} \right]_{nn} \leq  P_n,\\
& \forall n\in \{1\dots N_{t}\}, \notag
 \end{empheq}
 where $\mathbf p\in \mathbb{R}_{N_t}$ and $\mathbf x = \exp\left(\mathbf s{(t} )\right ) $.
\textcolor{black}{Problem $\mathcal{P}$ is a quadratic problem ($\mathrm{QP}$) \cite{convex_book} and can thus be solved to arbitrary accuracy using standard numerical methods\footnote{{Analytical methods to solve  problem $\mathcal{P}$ are beyond the scope of the present work. For more information, the reader is referred to \cite{convex_book}.}}.
} Subsequently, the solution of \eqref{eq: subgrad} is given as $\mathbf s (l+1) = \log\left(\mathbf p^*\right ) $, where $\mathbf p^* = \mathcal P\left(\mathbf p_{ant}, \mathbf x\right)$ is the optimal point of convex problem $\mathcal{P}$. To summarize the solution process, the per-antenna power constrained sum rate maximizing algorithm is given in Alg. \ref{Alg: MSR PAC}.
\begin{algorithm}
\SetAlgoLined 
\KwIn{(see Tab.\ref{tab: params}) $\{\mathbf w_{k}^{(0)}\}_{k=1}^{N_t} = \sqrt{P_{tot}/( N_t^{2})}\cdot\mathbf{1}_{N_t}$, $\mathbf p_{{ant}}$, $j=0$.  }
\KwOut{  $ \{\mathbf w_{k}^{\mathcal{SR}}\}_{k=1}^{N_t} $}
\Begin{
    \While{$\mathcal{SR}$ does not converge}{
    $j=j+1$\\
    \textbf{\textit{\uline{Step 1:}}}  Solve $r^{*}={Q}(\mathbf g_{(j)} ,  \mathbf p_{{ant}})$ to calculate $\{\mathbf w_{k}^{(j)}\}_{k=1}^{N_t}$. The input $\mathrm{SINR}$ targets $\mathbf g_{(j)} $ are given by the minimum $\mathrm{SINR}$ per group, i.e. $\gamma_i = \min_{i \in\mathcal{G}_k}\left\{\mathrm{SINR}_i\right\}, \forall i \in\mathcal{G}_k, k\in\{1\dots {N_t}\}$.\\
    \textbf{\textit{\uline{Step 2:}}} Initialize the sub-gradient search algorithm as: $\mathbf p ^{(j)} = \{p_{k}\}_{k=1}^{N_t} =\{||\mathbf w_{k}^{(j)}||^2_2\}_{k=1}^{N_t} $, $ \mathbf s^{(j)} = \{s_{k}^{}\}_{k=1}^{N_t} = \{\log  p_{k}\}_{k=1 }^{N_t} $, \textcolor{black}{$\{\mathbf v_{k}^{(j)}\}_{k=1}^{N_t} = \{\mathbf w_{k}^{(j)}/ \sqrt{p_{k}^{(j)}}\}_{k=1}^{N_t}$}.\\
    \textbf{\textit{\uline{Step 3:}}} Calculate $t_{max}$ iterations of the sub-gradient  power control algorithm, starting from $\mathbf s(0) = \mathbf s^{(j)} : $ \\
     \For{$t = 0 \dots t_{max}-1$}{ $\mathbf s{(t+1)} =\prod_{\mathbb P}\left[\mathbf s{(t)} - \delta(t) \cdot \mathbf{r}(t)\right]$ }
     $\mathbf s^{(j+1)}= \mathbf s(t_{max}-1)$,
      \\
     \textbf{\textit{\uline{Step 4:}}} Calculate the current throughput:
     $c^{*} = \mathcal{SR} \left(\mathbf p _{ant} \right)$ with
     $\{\mathbf w_{k}^{\mathcal{SR}}\}_{k=1}^{N_t} = \{\mathbf w_{k}^{(j+1)}\}_{k=1}^{N_t} = \{\mathbf v_{k}^{(j)}\exp(s^{(j+1)}_k)\}_{k=1}^{N_t}$
     }
}
 \label{Alg: MSR PAC}
\caption{ Sum-rate maximizing multigroup multicasting under per-antenna power constraints.}
\end{algorithm}

\begin{table}
\caption{Input Parameters for Alg. 1}
\centering
\begin{tabular}{l|l|l}
\textbf{Parameter}  &\textbf{Symbol} &\textbf{Value}\\\hline
 Sub-gradient iterations  & $t_{max}$ &  $1$   \\
 Sub-gradient initial value &$\delta(t)$&$0.4$\\
 Sub-gradient step &$\delta(t+1)$&$\delta(t)/2$\\
 Gaussian Randomizations&$N_{rand}$&$100$\\
 Per-antenna constraints &$ \mathbf p_{ant}$ & $P_{tot}/N_t\cdot\mathbf{1}_{N_t}$\\
User Noise variance& $\sigma_i^2$ &$1, \ \forall i \in\{1\dots N_{u}\}$
   \\\hline
\end{tabular}
\label{tab: params}
\end{table}
\subsection{Complexity \& Convergence Analysis }\label{sec: complexity2}
 An important discussion involves the complexity of the proposed algorithm. In \cite{Christopoulos2014ICC,Christopoulos2014}, the computational burden for an accurate approximate solution of  the per-antenna power minimization  problem $\mathcal {Q}$ (step 1 of Alg. 1) has been calculated. In summary, the relaxed power minimization is an semidefinite programming $\mathrm{(SDP)}$ instance with $ {N_t} $ matrix variables of $ N_t\times N_t $ dimensions and $N_{u}+N_t$ linear constraints. The present work relies on  the CVX tool \cite{convex_book} which calls  numerical  solvers such as SeDuMi to solve  semi-definite programs.  The interior point methods employed to solve this $\mathrm{SDP}$  require at most $\mathcal{O}({N_t}\log(1/\epsilon))  $ iterations, where $\epsilon $ is the desired numerical accuracy of the solver. Moreover, in each iteration not more than $\mathcal{O}({N_t^9 +N_{t}^{4} +N_{u}N_t^3})$ arithmetic operations will be performed.      The solver used \cite{convex_book} also exploits the specific structure of matrices hence the actual running time is reduced. Next,  a fixed number of iterations of the Gaussian randomization method is performed\cite{Luo2010}.  In each randomization, a linear problem ($\mathrm{LP}$) is solved    with a worst case complexity of  $\mathcal O( {N_t}^{3.5}\log(1/\epsilon_{1}) ) $ for an  $\epsilon_{1}-$optimal solution. The accuracy of the solution increases with the number of randomizations \cite{Karipidis2008,Sidiropoulos2006,Luo2010}.
    The remaining three steps of Alg. \ref{Alg: MSR PAC}
involve a closed form sub-gradient calculation as given in \cite{stanczak2009fundamentals} and the projection operation, which is a  real valued  least square problem under $N_t$ quadratic inequality $\mathrm{PAC}$s. Consequently, the asymptotic complexity of the derived algorithm is polynomial, dominated by the complexity of the $\mathrm{QoS}$ multigroup multicast problem under $\mathrm{PAC}$s.

The convergence of Alg. 1 is guaranteed given that the chosen step size satisfies the  conditions given in \cite{Kaliszan2012,stanczak2009fundamentals}, that is the diminishing step size. Herein,  $\delta(l+1) = \delta(l)/2$.
\textcolor{black}{What is more, in accordance to \cite{Kaliszan2012}, only a single iteration of the sub-gradient is performed in the numerical results (i.e. $t_{max} =1$)}.

\section{System Driven Optimization }\label{sec: SRM}
 \textcolor{black}{Constraints inspired by the inherent nature of satellite communications emanate the definition of novel optimization problems. The present section focuses on enabling  demanding in terms of availability satellite services. } Increased scepticism over spectrally efficient, aggressive frequency reuse, multibeam satellites stems from the effects of such configurations on the $\mathrm{SINR}$ distribution across the coverage.
%
In full frequency reuse scenarios, the useful signal power at the receiver is greatly reduced due to the intra-system interferences. Despite the throughput gains due to the increased user link bandwidth and  the adequate management of interferences by linear precoding,  the mean and variance of the  $\mathrm{SINR}$ distribution over the coverage area is generally reduced. This is the price paid for increasing the frequency reuse.  Naturally, this reduction in the average $\mathrm{SINR }$ will lead to a higher utilization of lower $\mathrm{MODCOD}$s and increase the probability of service unavailability over the coverage (outage probability). Retransmissions that incur in these outage instances, are bound to burden the system in terms of efficiency.    What is more, by acknowledging the multiuser satellite environment  (cf. Sec. \ref{sec: scheduling}),   these outage periods can potentially become comparable to the inherent long propagation delay of satcoms. Such a case will render the overall delay, as experienced by a user,  unacceptable.    As a result,   the probability of compromised users to experience long outage periods, needs to be considered in a system level. In this work, the introduction of minimum rate constraints over the entire coverage is proposed, as a means to guarantee in the physical layer design the stringent availability requirements typically accustomed in satcoms. The guarantee of a minimum level of service availability is introduced for the first time in a  $\max \mathrm{SR}$ multigroup multicast optimization.

\subsection{ Sum Rate Maximization under Minimum Rate Constraints }\label{sec: SRA}
 To provide high service availability, the  gains of the sum rate optimization can be traded-off in favor of a minimum guaranteed rate across the coverage.  This trade-off mostly depends on the minimum $\mathrm{MODCOD}$ supported by the $\mathrm{ACM}$\footnote{For instance in $\mathrm{DVB-S2X}$ under normal operation over a linearized channel, the most robust modulation and coding rate  can provide quasi error free communications (frame error probability lower than $10^{-5}$) for as low as  $-2.85$ dB of user $\mathrm{SINR}$, thus achieving a minimum spectral efficiency of $0.4348$ [bps/Hz]\cite{DVB_S2X}.
Beyond this value,  a service outage occurs.}.
Since an intermediate solution between the fairness and the $\max \mathrm{SR}$ goals is of high engineering interest, a novel optimization problem, namely the throughput maximization under  availability constraints, is proposed.  The innovation, aspired by operational requirements,  lies in the incorporation of minimum rate   constraints ($\mathrm{MRC}$s) in the $\mathrm{PAC}$  sum rate maximizing problem (equivalently minimum $\mathrm{SINR}$ constraints). Formally, the new optimization problem is defined as
 \begin{empheq}[box=\fbox]{align}
\mathcal{SRA:}&\max_{ \{\mathbf w_k \}_{k=1}^{N_t}}  \sum_{i=1}^{N_u} \log_2\left(1+\gamma_i \right) & \notag\\
\mbox{s. t.: }&  \gamma_i  = \min_{m\in \mathcal G_k}\frac{|\mathbf w_k^\dag \mathbf h_m|^2}{\sum_{l\neq k }^{N_t} |\mathbf w_l^\dag\mathbf h_m|^2+\sigma_m^2 }, &\label{const: SRA SINR}\\
&\forall i \in\mathcal{G}_k, k,l\in\{1\dots {N_t}\},\notag\\
 \text{and to: } & \left[\sum_{k=1}^{N_t}  \mathbf w_k\mathbf w_k^\dag  \right]_{nn}  \leq P_n, \label{eq: SRA PAC}\\%
 &\forall n\in \{1\dots N_{t}\},\notag\\
 \mbox{and to: } & \gamma_{i}  \geq \gamma_{min}, \  \forall i\in \{1\dots N_{u}\}.\label{const: SR MRC}
 \end{empheq}
In $\mathcal{SRA}$, the power allocation needs to account for the $\mathrm{MRC}$s, i.e. \eqref{const: SR MRC}. This is achieved by modifying the constraints of the sub-gradient search \cite{stanczak2009fundamentals}, as imposed via the projection of the current power vector onto the convex set of constraints. Therefore,  the additional constraint can be introduced in the projection method, since it does not affect the convexity of the formulation. Subsequently, to solve   $\mathcal{SRA}$ a new projection  that includes the minimum rate constraints is proposed. The new subset, that is the $\min \mathrm{SINR}$ constrained set, is a convex subset of the initially convex set. \textcolor{black}{The availability constrained projection reads as }
\begin{empheq}[box=\fbox]{align}
\mathcal{PA}: \min_{\mathbf p}  &||\mathbf p_{} -\mathbf x||^2_2& \notag\\
\mbox{subject to :} & \textcolor{black}{\frac{{p_k}|\mathbf v_k^\dag \mathbf h_i|^2}{\sum_{l\neq k }^{N_t} {p_l}| \mathbf v_l^\dag\mathbf h_i|^2 + \sigma^2_i}\geq \gamma_{min},}\label{eq: PA min SINR}\\
&\forall i \in\mathcal{G}_k, k,l\in\{1\dots {N_t}\},\notag\\
\mbox{and to :} & \left[\textcolor{black}{\sum_{k=1}^{N_t}  p_k\mathbf v_k  \mathbf v_k^\dag} \right]_{nn} \leq  P_n,\\
& \forall n\in \{1\dots N_{t}\}, \notag
 \end{empheq}
 \textcolor{black}{which is a convex optimization problem, that includes one additional linear constraint, i.e. \eqref{eq: PA min SINR}, over $\mathcal{P}$. \textcolor{black}{Provided that $\mathcal{SRA}$ is feasible, then \eqref{const: SR MRC} is satisfied and thus a  solution for $\mathcal{PA}$ always exists.} Similarly to $\mathcal{P}$, this problem can be solved using standard methods \cite{convex_book}.   }

Subsequently, the solution of $\mathcal{SRA}$ is derived following the steps of  Alg. \ref{Alg: MSR PAC} but with a modification in the  sub-gradient method (Step 3),  where the projection is calculated by solving problem $\mathcal{PA}$ instead of $\mathcal{P}$.
 As intuitively expected, the introduction of $\mathrm{MRC}$s
is bound to decrease the system throughput performance. However, this trade-off can be leveraged towards more favorable conditions, by considering other system aspects, as will be discussed in the following.
\subsection{Throughput Maximization via $\mathrm{MODCOD}$ Awareness}
A modulation constrained practical system employs higher order  modulations to increase its rate with respect to the useful signal power. The strictly increasing logarithmic cost functions describe communications based on Gaussian alphabets and provide the Shannon upper bound of the system spectral efficiency. Therefore, the sum rate maximization problems solved hitherto fail to account for the modulation constrained throughput performance of practical systems. The complication lies in the analytically intractable, at least  by the methods considered herein, nature of a step cost function. In the present section, an attempt to leverage this cost function in favor of the system throughput performance is presented. In more detail, benefiting from the  finite granularity of the rate function \eqref{eq: throughput} over the achieved $\mathrm{SINR}$, an extra system level optimization can be defined as
 \begin{empheq}[box=\fbox]{align}
\mathcal{SRM:}&\max_{ \{\mathbf w_k \}_{k=1}^{N_t}}  \sum_{i=1}^{N_u} f_{\mathrm{DVB-S2X}}\left(\gamma_i , \mathbf t \right) & \notag\\
\mbox{s. t.: }&  \gamma_i  = \min_{m\in \mathcal G_k}\frac{|\mathbf w_k^\dag \mathbf h_m|^2}{\sum_{l\neq k }^{N_t} |\mathbf w_l^\dag\mathbf h_m|^2+\sigma_m^2 }, &\label{const: SRM SINR}\\
&\forall i \in\mathcal{G}_k, k,l\in\{1\dots N_t\},\notag\\
 \text{and to: } & \left[\sum_{k=1}^{N_t}  \mathbf w_k\mathbf w_k^\dag  \right]_{nn}  \leq P_n, \label{eq: SRM PAC}\\%
 &\forall n\in \{1\dots N_{t}\},\notag\\
 \mbox{and to: } & \gamma_{i}  \geq \gamma_{min}, \  \forall i\in \{1\dots N_{u}\},\label{const: SR MRC2}
 \end{empheq}
where $f_{\mathrm{DVB-S2X}}(\cdot,\cdot)$ is the finite granularity step function defined in \eqref{eq: throughput}. 
The realization of a non-strictly increasing cost function   inhibits the application of gradient based solutions and necessitates a different solution process. 
 To provide a solution for this elaborate -yet of high practical value- problem, a heuristic iterative  algorithm is
proposed.
 More specifically,   Alg. \ref{Alg: MSR PAC DISCR} receives as input the availability constrained precoders $\{\mathbf w_{k}^{\mathcal{SRA}}\}_{k=1}^{N_t} $ calculated as described in Sec. \ref{sec: SRA}, and calculates an initial $\mathrm{SINR}$ distribution.   Then, it derives new precoding vectors under minimum $\mathrm{SINR}$ constraints given by the closest lower threshold of the worst user in each group, according to the discrete throughput function. Therefore, the resulting system throughput is not decreased while power is saved. This power can now be redistributed.  Also, in this manner, the solution guarantees a minimum system availability. Following this step, an ordering of the groups takes place, in terms of minimum required power to increase each group to the next threshold target. For this, the power minimization problem is executed for each group. Next,  each of the available groups, starting from the group that requires the least power,  is sequentially given a higher  target.   With the new targets, the power minimization problem is again solved. This constitutes a  feasibility optimization check.  If the required power satisfies the per antenna constraints, then these precoders are kept. Otherwise the current group is given its previous feasible $\mathrm{SINR}$  target and the search proceeds to the next group.

\textcolor{black}{\textit{Remark:} A further improved solution can be attained when dropping the constraint of a single step increase per group. Herein, such a consideration is avoided for complexity reasons. Since each of the ${N_t}$ groups can take at most ${N_m}$ possible $\mathrm{SINR}$ values, where ${N_m}$ denotes the number of $\mathrm{MODCOD}$s, by allowing each group to increase more than one step, the number of possible combinations can be as much as $(N_m)^{N_t}$. As a result, the complexity of the optimal solution found by searching the full space of possible solutions, grows exponentially with the number of groups. In the present work, the high number of threshold values for $f_{\mathrm{DVB-S2X}}$ prohibits such considerations.}

The summary of this algorithm is given in Alg. \ref{Alg: MSR PAC DISCR}. Since it is an iterative algorithm over the number of available groups, convergence is guaranteed. Also, since it receives as input the $\mathcal{SRA}$ solution, its complexity is dominated by the complexity of Alg. \ref{Alg: MSR PAC}, as described in Sec. \ref{sec: complexity2}.

\begin{algorithm}
 \SetAlgoLined 
 \KwIn{$  \mathbf H,P_{tot}, \sigma_i^2 \ \forall i \in\{1\dots N_t\},  \{\mathbf w_{k}^{(0)}\}_{k=1}^{N_t} =  \{\mathbf w_{k}^{\mathcal{SRA}}\}_{k=1}^{N_t}  $, $r^{(0)} $,  $\textcolor{black}{\gamma_{min}}$ }
 \KwOut{   $ \{\mathbf w_{k}^{out}\}_{k=1}^{N_t} $ }
 \Begin{ $j=0$; $q=1$; $ \{\mathbf w_{k}^{out}\}_{k=1}^{N_t} = \{\mathbf w_{k}^{(0)}\}_{k=1}^{N_t}  $;\\
 \textbf{\textit{\uline{Step 1: }}} Solve  \textcolor{black}{$r^{*, (0)}=\mathcal{Q}(\mathbf g^{(0)} ,  \mathbf p_{ant})$} to calculate $\{\mathbf w_{k}^{\mathcal{Q},(0)}\}_{k=1}^{N_t}$. The input $\mathrm{SINR}$ targets are given by the minimum threshold $\mathrm{SINR}$ per group, i.e. $\mathbf g^{(0)}: \gamma_i = \left\lfloor \min_{m \in\mathcal{G}_k}\left\{\mathrm{SINR}_m\right\}\right\rfloor_{\mathbf t}, \forall i,  m \in\mathcal{G}_k, k =1,\dots ,N_t$.\\
\For{ $j=1\dots {N_t}$ }{
\textbf{\textit{\uline{Step 2: }}} Solve  $r^{*,(j)}= \mathcal{Q}(\mathbf g^{(j)} , \mathbf p)$ to calculate $\{\mathbf w_{k}^{\mathcal{Q},(j)}\}_{k=1}^{N_t}$.
 The targets of the current $j$-th group are increased by one level: $\gamma_i = \left\lceil \min_{m \in\mathcal{G}_j}\left\{\mathrm{SINR}_m\right\}\right\rceil_{\mathbf t}, \forall i \in\mathcal{G}_j $;\\ Order the groups in terms of increasing $r^{*,(j)}$.\\
}
 \While{ $r^{*,(q)}<1$ }{{\textbf{\textit{\uline{Step 3: }}}} For each group, in a sequence ordered by the previous step, increase the target by one level; \\ Solve  $ r^{*,(q)}= \mathcal{Q}(\mathbf g^{(q)} , \mathbf p)$ with input targets from the previous iteration: $\mathbf g^{(q)}= \mathbf g^{(q-1)}$; $q=q+1$}
 $ \{\mathbf w_{k}^{out}\}_{k=1}^{N_t} =\{\mathbf w_{k}^{\mathcal{Q},(q)}\}_{k=1}^{N_t}$
}
\label{Alg: MSR PAC DISCR}
 \caption{Discretized sum rate maximization.}
\end{algorithm}

\section{User Scheduling }\label{sec: scheduling}
\begin{figure}[!htp]
 \centering
 \includegraphics[width=0.8\columnwidth]{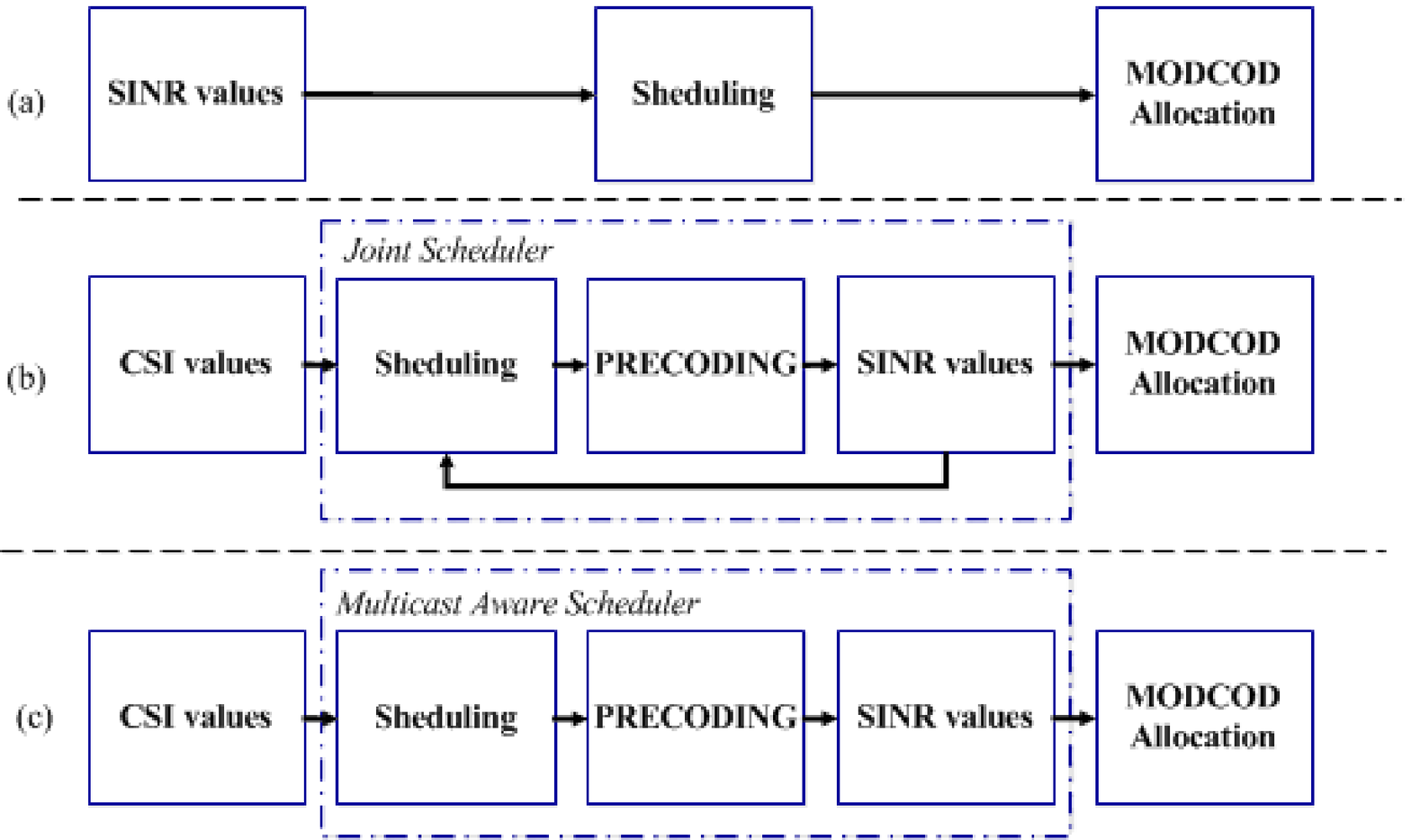}\\
  \caption{\textcolor{black}{Scheduling over satellite: (a) Conventional $\mathrm{DVB-S2}$ (b) Optimal joint precoding and scheduling (c) Proposed multicast-aware heuristic scheduling.}}
\label{fig: scheduling}
\end{figure}

 Multibeam satellite systems typically  cover vast areas by a single satellite illuminating  a large pool of users requesting service. Therefore, a satcom system  operates in  a  large multiuser  environment. In  current satcom standards,  user scheduling is  based  on the traffic demand and channel quality  \cite{DVB_S2_standard}. Thus   $\mathrm{DVB-S2}$ schedules    relatively similar in terms of $\mathrm{SINR}$ users in the same frame and a specific link layer mode (assuming  $\mathrm{ACM}$) is employed to serve them. A diagram with the necessary operations performed at the transmitter is illustrated in Fig. \ref{fig: scheduling} (a) for conventional systems.
In aggressive resource reuse transmitters that employ precoding, scheduling policies can be based on the principles of $\mathrm{MU-MIMO}$ communications.  The inherent difference with conventional systems is that the $\mathrm{CSI}$ for each user is now an $N_t$ dimensional vector rather than a single $\mathrm{SINR}$ value. In the parlance of $\mathrm{MU-MIMO}$ communications the level of similarity between the users  can be measured in terms of orthogonality of the complex vector channels.   To maximize the similarity of two vectors, one needs to maximize their projection, that is the  dot product of the two vectors. On the contrary, to maximize their orthogonality, the projection needs to be minimized. As it will be shown hereafter, by accounting the vector $\mathrm{CSI}$ in the scheduling process, the multiuser gains can be exploited towards further maximizing the system throughput performance.

 Inspired by the multigroup multicast nature of the frame-based precoding problem, a multicast-aware user scheduling policy is developed in the present section.  In   the frame-based precoding methods presented in the previous sections, a precoding design over a randomly defined group of users is assumed. Since all co-scheduled  users are served by the link layer mode imposed by the worst user in each group,  significant performance losses from a system design perspective will be realized by this random user grouping. Acknowledging that $\mathrm{CSI}$ is readily  available at the transmit side, since it  is a requisite for the application of interference management,   the optimization of the system in any required sense can be achieved by advanced scheduling methods. These methods, as shown in Fig. \ref{fig: scheduling} (b) and (c) are based on the exact $\mathrm{CSI}$.  Imperfect $\mathrm{CSI}$ assumptions shall be tackled in future extensions of this work.

The most intrinsic attribute of a joint scheduling and precoding design lies in the coupled nature of the two designs. Since precoding drastically affects the useful signal power at the receive side, the relation between $\mathrm{CSI}$ and $\mathrm{SINR}$ is not straightforward.  The block diagram in Fig. \ref{fig: scheduling} (b),  presents an optimal joint scheduler. This module jointly performs precoding and scheduling by feeding  the output of the precoder back to the scheduler. Based on an initial user scheduling, a precoding matrix calculated by the methods of Sec. \ref{sec: SRM},  can be applied. Then, the resulting $\mathrm{SINR}$  value needs to be fed back to the scheduler where a new schedule is then re-calculated. Based on this schedule, a new precoding matrix needs to be calculated and applied thus leading to a potentially different $\mathrm{SINR}$ distribution. Clearly, this procedure needs to be  performed until all the possible combinations of users are examined. Thus, the implementation complexity of such a technique is prohibitive for the system dimensions examined herein. A reduction of the system dimensions, on the other hand, reduces the averaging accuracy and renders the results inaccurate from a system design perspective. Therefore, the optimal user scheduling policy will not be considered for the purposes of this work.

 As described in the previous paragraph, precoding is affected by scheduling and vice versa. To the end of providing a low complexity solution to this causality dilemma, a multicast-aware approach is illustrated in Fig. \ref{fig: scheduling} (c). Based on this concept, an advanced low complexity $\mathrm{CSI}$  based scheduling method that does not require knowledge of the resulting SINR, is developed.  The key step in the proposed method  lies in measuring the similarity between  user channels, given the readily available $\mathrm{CSI}$. The underlying intuition  is that users scheduled in the same frame should have co-linear (i.e. similar) channels since they need to receive the same set of symbols (i.e. frame).  On the contrary,  interfering users, scheduled in adjacent synchronous frames,   should be  orthogonal to minimize interferences \cite{Yoo2006b}.  The multicast-aware user scheduling  algorithm, presented in detail in Alg. \ref{alg: MAUS}, is a low complexity heuristic iterative algorithm that allocates orthogonal users in different frames and simultaneously parallel users with similar channels in the same frame. In more detail,  this two step algorithm operates as follows. In the first step of the process,   one user per group is allocated according to the semi-orthogonality criteria originally proposed in \cite{Yoo2006b}. This semi-orthogonality criterion was originally derived for zero-forcing $\mathrm{ZF}$ precoding, in order to find the users with the minimum interferences. This approach is adopted for the first step of the proposed algorithm, since the goal is to  allocate non-interfering users in different groups. Next, a novel second step provides the multicast awareness of the herein proposed algorithm.    In Step 2, for each of the groups sequentially, the most parallel users to the previously selected user are scheduled in the same frame. Subsequently, the similarity of the co-group channels is maximized.

\begin{algorithm}
\SetAlgoLined
\KwIn{$\mathbf H $}
\KwOut{User allocation sets $\mathcal{I}$ }
\Begin{
    \uline{\emph{Step 1:}}
     $\forall \ l = 1, 2 \dots N_{t} $ allocate  semi-orthogonal users to different groups. Let $\mathcal{I} = \emptyset $ denote the index set of users allocated to groups,  $\mathcal{J}=\{1,\dots N_{u} \} - \{ \mathcal{I}\} $ the set of unprocessed users and \textcolor{black}{$g_{(1)}=  \max_{k}||\mathbf h_{k}||_2$ }  \\
    \While{$|\mathcal{I}| < N_{t} $ }
    {
    \ForAll{ $m \in\mathcal{J}$, $l =1\dots N_t$ }
    {\textcolor{black}{$\mathbf g_{m}^\dag = \mathbf h_{m}^\dag\left(\mathbf{I}_{N_{t}}-\sum_{q=1}^{l} \frac{\mathbf{g}_{(q)} \mathbf{g}_{(q)}^\dag}{||\mathbf{g}_{(q)}||_2^2} \right)$}  calculate the orthogonal component (rejection) of each unprocessed user's channel, onto the subspace spanned by the previously selected users.}
    Select the most orthogonal user to be allocated to the $l$-th group:
    $\mathcal{G}_l = \arg \max _{m} ||\mathbf g_{m}||_{2} $ ,
    $\mathbf g_{(l)} = \mathbf g_{\mathcal{G}_l}$ and update the user allocation sets $\mathcal{I} = \mathcal{I} \cup\{\mathcal{G}_l\}$,  $\mathcal{J}=\mathcal{J} - \{ \mathcal{G}_l\}$\\
    }
    \uline{\emph{Step 2:}} for each group select the most parallel users.\\
    \For{$l=1\dots N_{t}$}
        {
        \While{$|\mathcal{G}_l| < \rho$}
            {
            \ForAll{ $m \in \mathcal{J}$}
                {
    $ \mathbf u_m\ = \mathbf h_m^\dag\frac{\mathbf h_{j}\mathbf h_{j}^\dag}{\textcolor{black}{||\mathbf h_{j}^\dag||_{2}^2}}, j= [\mathcal{G}_l ]_{1}$; calculate the projection of each users channel, onto the first user of each group.
    Select the  user that is most parallel to the first user of each group.
    $\pi_{l} = \arg \max _{m} \{||\mathbf u_{m}||_{2}\} $  and update the user allocation sets $\mathcal{G}_l = \mathcal{G}_l\cup\{\pi_l\}$, $\mathcal{I} = \mathcal{I} \cup\{\mathcal{G}_l\}$,
    $\mathcal{J}=\mathcal{J} - \{ \mathcal{G}_l\}$\\
                }
            }
        }
    }
\label{alg: MAUS}
\caption{Multicast-Aware User Scheduling Algorithm}
\end{algorithm}
\section{ Performance Evaluation \& Applications} \label{sec: performance}
Based on the simulation model defined in \cite{satnex}, the performance  of a  full frequency reuse, broadband multibeam satellite that employs frame-based precoding, is compared to conventional four color reuse configurations. \textcolor{black}{Since by the term user, a individual receive terminal is implied and  the terms frame, beam and group are effectively equivalent in the scenario under study, the total number of users considered over the entire coverage can be found by multiplying the users per frame with the number of beams. }The average user throughput given  by \eqref{eq: throughput} is calculated  to quantify the potential gains of frame-based precoding.
\begin{table}
\caption{Link Budget  Parameters}
\centering
\begin{tabular}{l|l}
\textbf{Parameter}  & \textbf{Value}\\\hline
 Frequency Band  &  Ka (20~GHz)   \\
 User terminal clear sky temp,   $T_{cs}$  & 235.3K\\
 User Link Bandwidth,   \(B_u\) & 500~MHz\\
 Output Back Off, $\mathrm{OBO}$&5~dB\\
 On board Power, $P_{tot}$ & 50~dBW\\
Roll off, $\alpha$ & 0.20 \\
 User terminal antenna  Gain,  $G_R$& 40.7~dBi \\
  Multibeam Antenna Gain, $G_{ij}$ &Ref: \  \cite{satnex}
  \\\hline
\end{tabular}
\label{tab: simulation params}
\end{table}
The rate and $\mathrm{SINR}$ distributions over the coverage before and after precoding are also investigated. Moreover, the sensitivity of all discussed methods to an increasing number of users per frame is presented.   The simulation setup is described in Sec. \ref{sec: channel}. For accurate averaging, 100 users per beam are considered uniformly distributed across the coverage area illustrated in Fig. \ref{fig: CA}. The average user throughput $R_{avg}$, as given via \eqref{eq: throughput}, is also averaged over all transmissions required to serve the initial pool of users. This consideration provides a fair comparison when user scheduling methods are considered\footnote{Serving less users than the available for selection would drastically improve the results but not in a fair manner from a system design perspective, since this would imply that some users are denied service for an infinite time.}. The link budget parameters considered follow the recommendations of \cite{satnex} and are summarized in Tab. \ref{tab: simulation params}. \textcolor{black}{The minimum $\mathrm{SINR}$ value $\gamma_{\min}$ considered herein is $-2.85$~dB, corresponding to the minimum value supported by the normal frame operation of the most recent satcom standards \cite{DVB_S2X}. Operation in even lower values is bound to increase the reported gains, since a relaxation in the added availability constraint allows for higher flexibility and thus sum-rate gains. }
\subsection{Throughput performance}\label{sec: throughput no sched}
\begin{figure}[!htp]
 \centering
 \includegraphics[width=0.75\columnwidth]{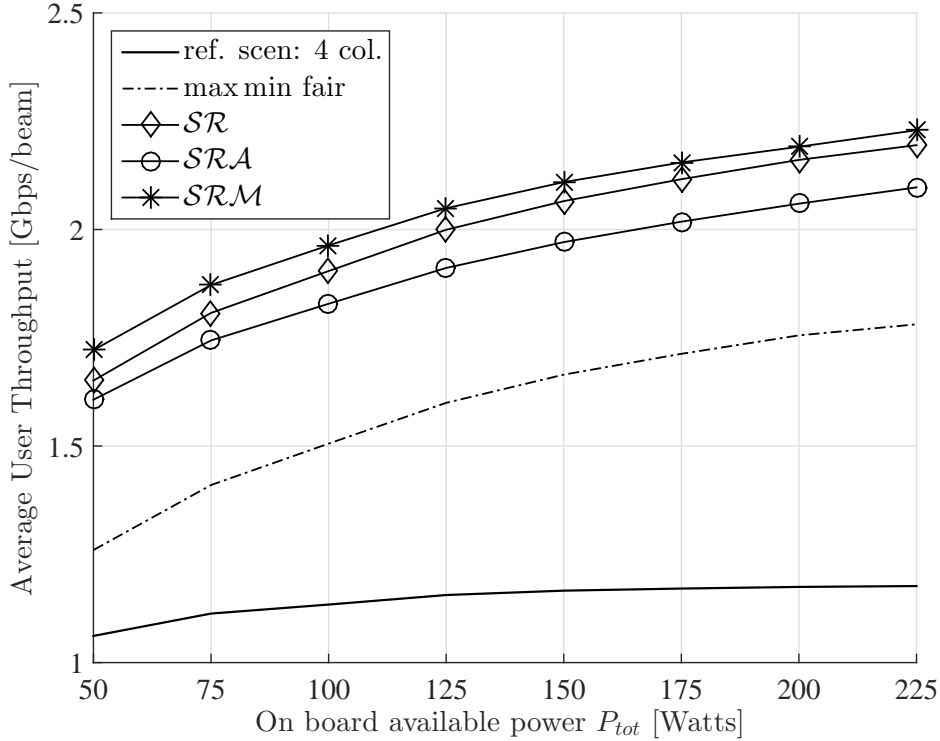}\\
  \caption{\textcolor{black}{Average user throughput versus on board available transmit power, for 2 users per frame.}}
\label{fig: power1}
\end{figure}

\textcolor{black}{The validity of the heuristic sum-rate maximization algorithm is established by comparing the performance of the herein proposed precoders with the optimal in a $\max-\min$ fair sense, solutions of \cite{Christopoulos2014}. The throughput versus availability tradeoff  between the two formulations will  also be exhibited in the following. } In Fig. \ref{fig: power1}, the average user throughput of the considered multibeam satellite is plotted  versus an increasing total on board available  power, in [Gbps/beam]. Two users per frame are considered, i.e. $ \rho = 2$. Clearly, the proposed precoding designs outperform existing approaches.   The  $ \mathcal{SR}$ problem achieves more than 30\% gains over the $\max \min$ fair solutions of \cite{Christopoulos2014ICC,Christopoulos2014}. These gains are reduced when the $\max \mathrm{SR}$ under $ \mathrm{MRC}$s is considered, i.e. $ \mathcal{SRA}$. This is the price paid for guaranteeing service availability over the coverage. Finally, the maximum gains are observed when the modulation aware $\max \mathrm{SR}$ precoding, i.e. $ \mathcal{SRM}$ is employed, which also guarantees service availability. Consequently, the best performance is noted for  $\mathcal{SRM}$ with more than 30\% of gains over the $\max \min$ fair formulation of \cite{Christopoulos2014} and as much as 100\% gains over  conventional systems in the high power region, for 2 users per frame.

\begin{figure}[!htp]
 \centering
 \includegraphics[width=0.75\columnwidth]{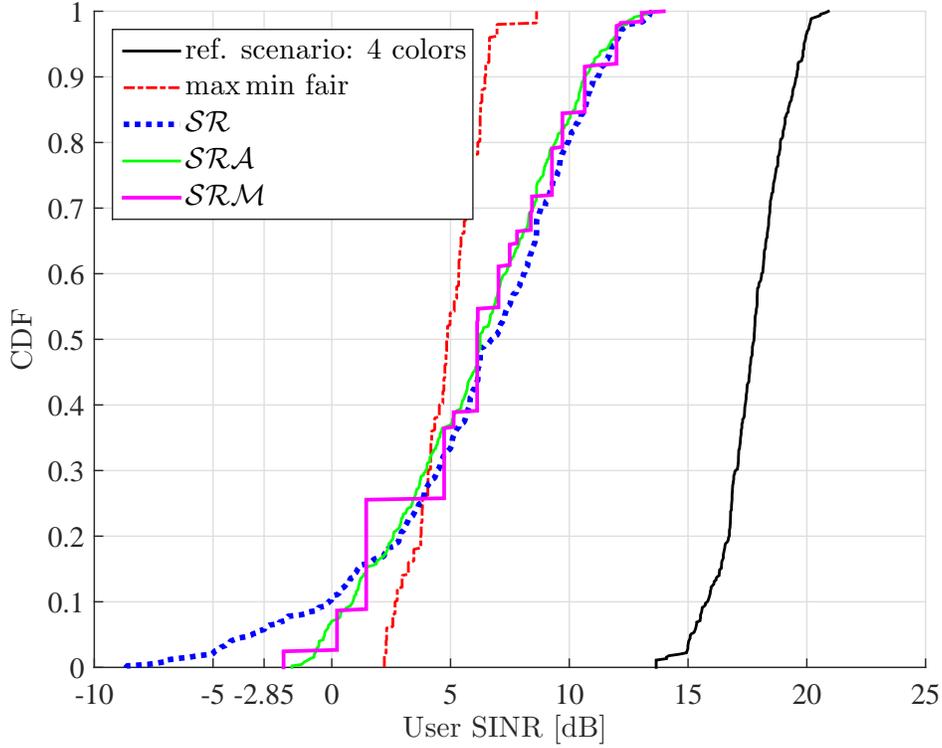}\\
  \caption{ \textcolor{black}{CDF of user   $\mathrm{SINR}$ over the coverage, for 2 users per frame.}}
\label{fig: SINR CDF}
\end{figure}

For the same simulation setting, the cumulative distribution functions $\mathrm{(CDF}$s) of the  $\mathrm{SINR}$s over the coverage area is given in Fig. \ref{fig: SINR CDF}. Clearly, conventional systems  achieve higher $\mathrm{SINR}$s  by the means of the fractional frequency reuse. This value is  around $17$ dB, in line with the results of \cite{satnex}. However, this does not necessarily translate to system throughput performance. To guarantee increased $\mathrm{SINR}$s, the frequency allocated per user is four times reduced. On the other hand, aggressive frequency reuse reduces the average $\mathrm{SINR}$ values and increases its variance, as seen in Fig. \ref{fig: SINR CDF}. This, however, allows for more efficient resource utilization and consequently higher throughput, as seen before in Fig. \ref{fig: power1}. Moreover, the superiority of the $ \max \mathrm{SR}$ techniques proposed herein, over the fair solutions is also evident. Amongst these methods, the best one is $\mathcal{SRM}$ as already shown.

\textcolor{black}{The benefits of $\mathcal{SRA}$ over $\mathcal{SR}$ are clear in Fig. 6, where the $\mathrm{SINR}$ $\mathrm{CDF}$ of all methods is presented. Clearly, $\mathcal{SRA}$ guarantees a minimum $\mathrm{SINR}$ of -2.85~dB but attains $\mathrm{SINR}$s  higher than 2~dB with less probability than $\mathcal{SR}$. Nevertheless,  $\mathcal{SRA}$ can be regarded as a middle step towards advancing to the more elaborate, $\mathcal{SRM}$ algorithm. Since $\mathcal{SRM}$  includes the same  availability constraints as $\mathcal{SRA}$,   identical availability gains are noted for both methods. However, $\mathcal{SRM}$ exploits the granular nature of the spectral efficiency function towards achieving $\mathrm{SINR}$s higher than $\mathcal{SR}$. }  In Fig. \ref{fig: SINR CDF}, it is clear that the proposed optimization manages to adapt each user's $\mathrm{SINR}$ to the throughput function, since the  $\mathrm{SINR}$ distribution follows the granular spectral efficiency function. Users have  $\mathrm{SINR}$ values in between the DVB-S2X thresholds with very low probability. This insightful result justifies the increased gains of $\mathcal{SRM}$, even for guaranteed availability. An additional observation from Fig. 6 is that 40\% of the users operate utilizing the first four available $\mathrm{MODCOD}$s.

\begin{figure}[!htp]
 \centering
 \includegraphics[width=0.75\columnwidth]{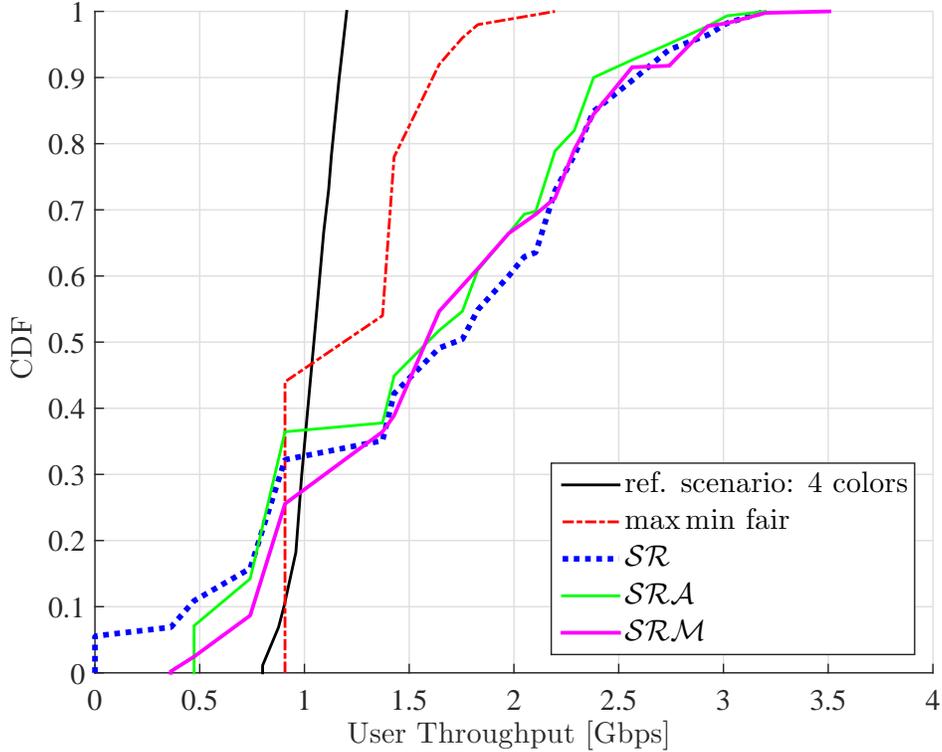}\\
  \caption{\textcolor{black}{Per-user throughput CDF, for 2 users per frame.}}
\label{fig: rate1}
\end{figure}

Moreover, Fig. \ref{fig: rate1} provides the rate $\mathrm{CDF}$s of the conventional and the $\max \min $ fair systems and exhibits the very low variance of their receive $\mathrm{SINR}$. On the contrary, $\mathcal{SR}$  achieves very high rates but also drives some users to the unavailability region. A 5\% outage probability is noted for this precoding scheme. This is not the case for the  $\mathcal{SRA}$ and  $\mathcal{SRM}$ problems,  which guarantee at least 0.3 Gbps to all users.

\begin{figure}[!htp]
 \centering
 \includegraphics[width=0.75\columnwidth]{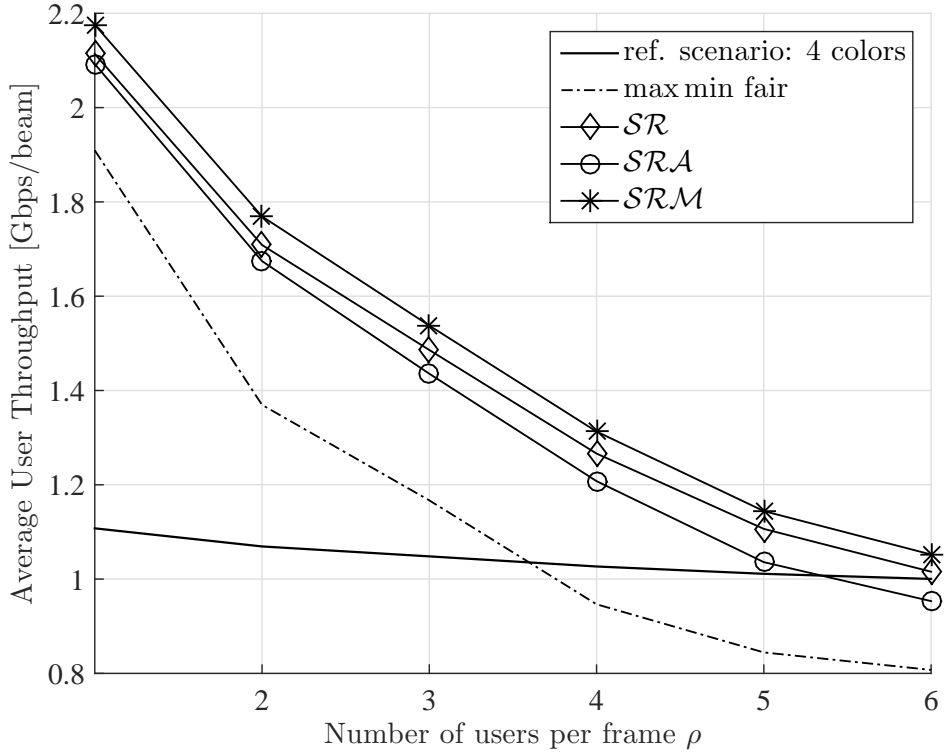}\\
  \caption{\textcolor{black}{Average user throughput versus the number of users per frame.}}
\label{fig: user1}
\end{figure}

An important issue is the performance of the developed methods with respect to an increasing number of users per frame. As presented in Fig. \ref{fig: user1},  $\mathcal{SRM}$ manages to provide more than 30\% of gains for $\rho = 3$ users per frame.  Both the conventional and the proposed systems suffer from an increase in the number of users per frame, since the worst user defines the $\mathrm{MODCOD}$ for all users. For conventional systems, this degradation is negligible when compared to the frame based precoding systems. The performance degradation when a precoding vector is matched to more than one channels is expected. As initially proven in \cite{Sidiropoulos2006}, when advancing from unicasting to multicasting, the precoding problem becomes $\mathrm{NP}$-hard. Added to that,  when more users are grouped together, then the chances are that one of them will be  compromised and thus constrain the performance of all other users. This  observation further justifies the results of Fig. 8.  Nevertheless, in the same figure, positive gains over the conventional systems are reported even for 6 users per frame unlike all other state of the art techniques. These results are given for a nominal on board available power of $50$ Watts. It should be noted that  performance in the results presented hitherto is compromised by the random user scheduling since users with very different $ \mathrm{SINR }$s are co-scheduled and thus constrained by the performance of the worst user.
\subsection{ Example}
\begin{figure}[!htp]
 \centering
 \includegraphics[width=0.75\columnwidth]{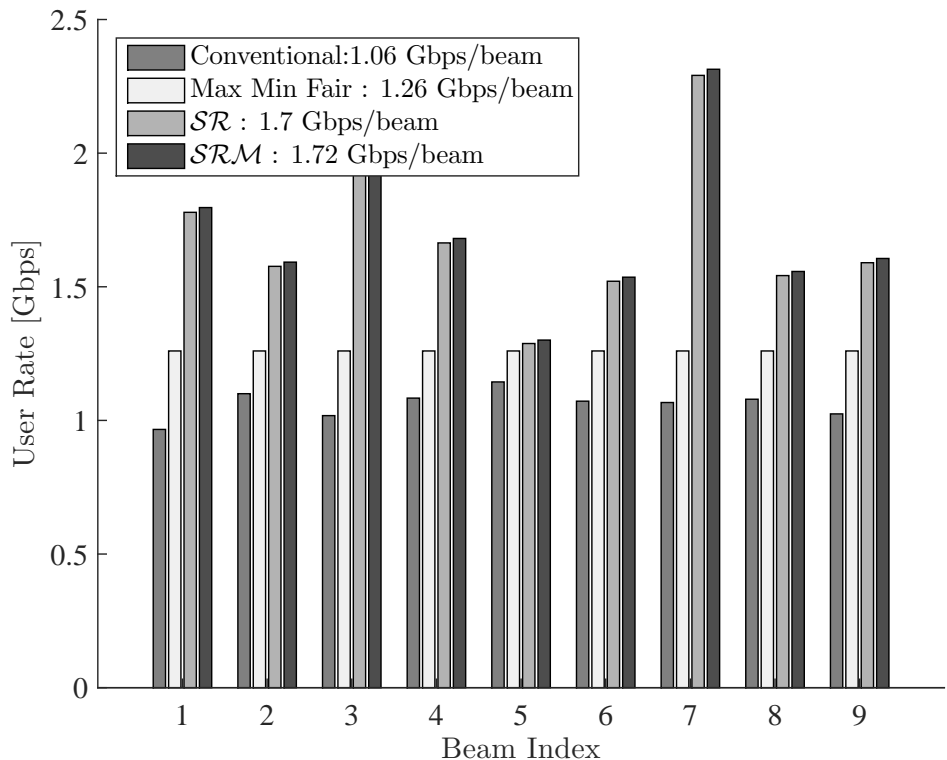}\\
  \caption{\textcolor{black}{Per-user achievable rate in each beam, for different designs.}  }
\label{fig: paradigm}
\end{figure}

To the end of gaining insights on  the $\max \mathrm{SR }$ optimization, a small scale example   is presented.  Let us  assume $2$ users per frame (i.e. $\rho = 2$).
 The individual throughput of each user is plotted in  Fig. \ref{fig: paradigm} for the discussed methods. The per beam average throughput  is given in the legend of the figure for each method respectively. In the conventional system,  variance in the rates between the groups is noted. This results to an average user throughput equal to  $1.06$~Gbps/beam. By the fair optimization of \cite{Christopoulos2014}  $1.26$ Gbps/beam of are attained, while  the minimum rates are balanced among the groups. More importantly, the sum rate maximizing optimization  reduces the rate allocated to the users in beam $5$ and increases all other users. Thus, the system throughput  is increased to just over $1.6$ Gbps/beam.  Finally,  the modulation aware optimization builds upon the sum rate maximization, adapts the power allocation to the modulation constrained performances and allocates to each user equal or better rates. Consequently, it outperforms all other techniques leading to $R_{avg} = 1.72$~Gbps/beam.
\subsection{User scheduling}
 The present section presents results when  the multicast-aware user scheduling algorithm is employed. In Fig. \ref{fig: power2}, the performance of the algorithm for $\rho = 2$ users per group is given versus an increasing on-board power budget. In this figure, approximately 25\% of improvement the random scheduling of Sec. \ref{sec: throughput no sched}  is noted. Furthermore, in Fig. \ref{fig: user2},  results for an increasing number of users per frame and for a nominal on board available power of $50$ Watts, are given.  The performance of $\mathcal{SRM}$ without scheduling as presented in Fig. \ref{fig: user1}, is also given for comparison.  From the results of Fig. \ref{fig: user2},   it is clear that by employing user scheduling, the degradation of the system performance with respect to an increasing number of users per group is significantly improved. The same initial group of users as before is employed regardless of the frame size, excluding a small rounding error cut off\footnote{For instance, when 3 users per frame are assumed, the total number of users served is reduced to 891. This does not affect the presented results, since they are averaged over the total number of users served.}. The most important result is that by employing multicast-aware user scheduling methods, more than 30\% of gains can be gleaned over conventional systems for as much as 7 users per frame. Also, even 13 users per frame can be accommodated in a frame with positive gains over conventional frequency reuse payload configurations.
\begin{figure}[!htp]
 \centering
 \includegraphics[width=0.75\columnwidth]{power_TWCOM_MAUS_r2}\\
  \caption{\textcolor{black}{Average user throughput versus on board available transmit power, for 2 users per frame, when scheduling is employed.}}
\label{fig: power2}
\end{figure}
\begin{figure}[!htp]
 \centering
 \includegraphics[width=0.75\columnwidth]{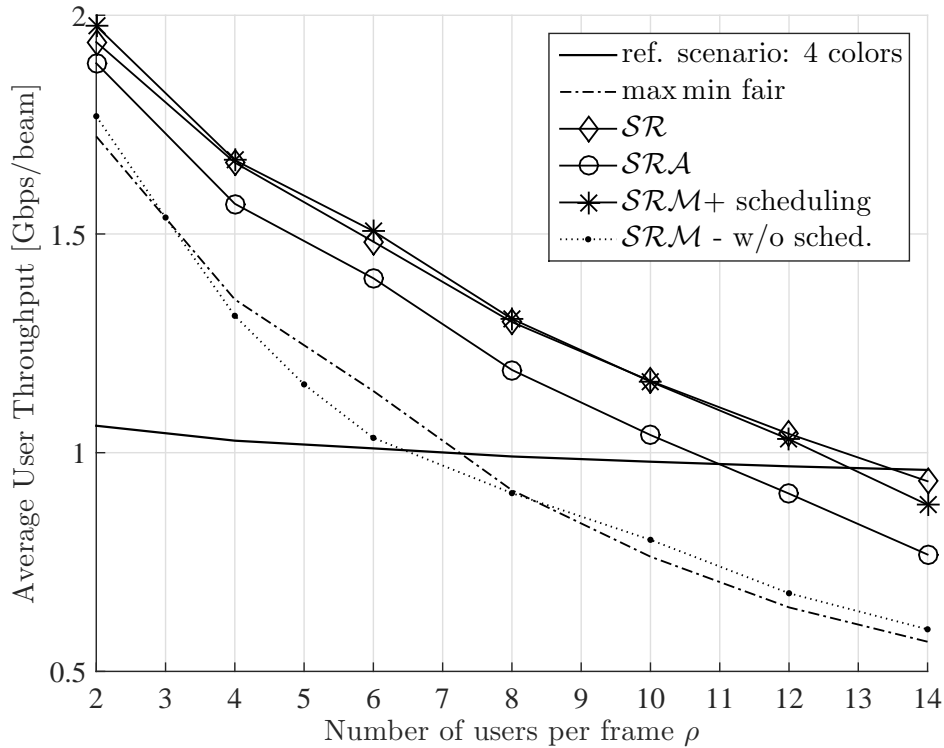}\\
  \caption{\textcolor{black}{Per beam throughput with respect to an increasing number of users per frame.}}
\label{fig: user2}
\end{figure}
Finally, to exhibit the dependence of the performance with respect to the available for selection user pool, in Fig. \ref{fig: user pool}, the average user throughput for three users per frame with respect to an increasing user pool is plotted. Almost 20\% gains are noticed when doubling the user pool.
Clearly, the potential of user scheduling is even higher in larger multiuser settings.
\begin{figure}[!htp]
 \centering
 \includegraphics[width=0.75\columnwidth]{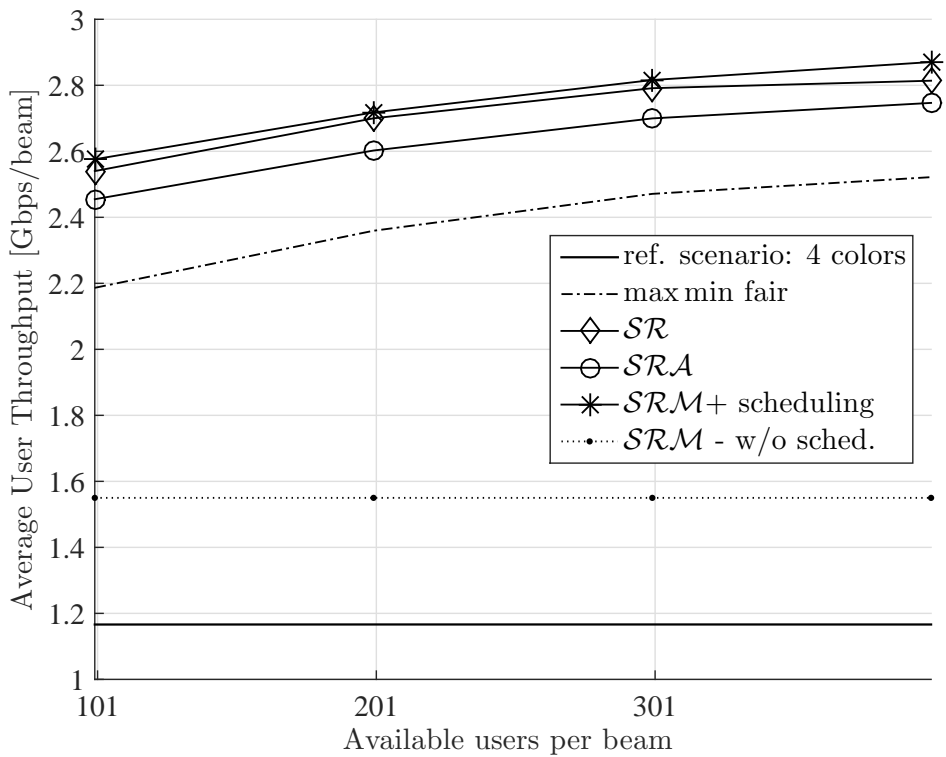}\\
  \caption{\textcolor{black}{Average  throughput with respect to an increasing number of available for selection users, for 3 users per frame, when scheduling is employed. }}
\label{fig: user pool}
\end{figure}
\section{Conclusions} \label{sec: conclusions}
In the present work, full frequency reuse configurations enabled by frame-based linear precoding are proposed for the optimization of broadband multibeam satellite systems in terms of throughput performance.
In this direction, a sum rate optimal  frame-based precoding design  under per-antenna power constraints is derived.
To satisfy the highly demanding in terms of availability satcoms requirements, while maintaining high gains over conventional systems, the optimization is extended to account for minimum rate constraints as well as the modulation constrained throughput performance of the system. Finally, to glean the satellite multiuser diversity gains, user scheduling methods adapted to the novel system design are derived. In summary, the gains from frame-based precoding combined with multicast-aware user scheduling are more than 30\%    in terms of throughput performance, for $7$ users per frame, over conventional system configurations.    These gains are achieved without loss in the outage performance of the system. Also   up to 13 users per frame can be accommodated with throughput performance similar to that of the conventional systems. 

Future extensions of this work include a robust frame-based precoding design  to cope with $\mathrm{CSI }$ imperfections as well as studies to counteract the non-linearities of the  satellite channel.

\bibliographystyle{IEEEtran}

\begin{IEEEbiography}[{\includegraphics[width=1.0in,height=1.25in,clip,keepaspectratio]{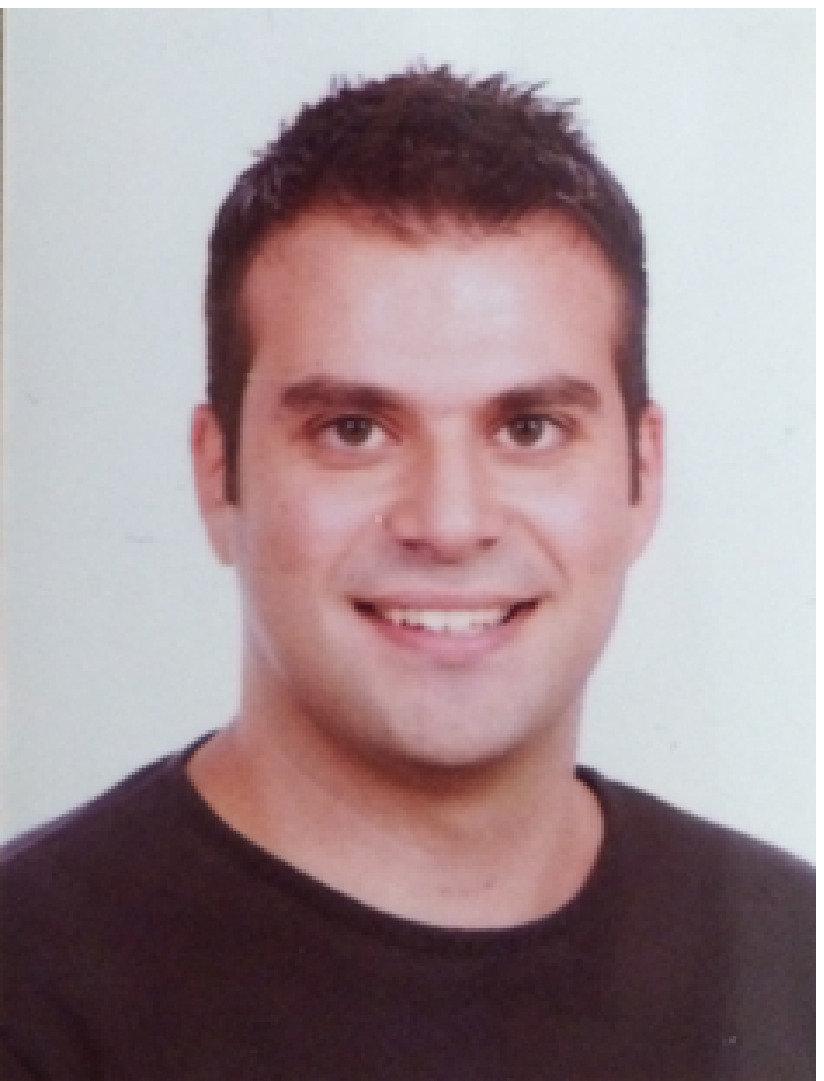}}]{\textbf{Dimitrios Christopoulos (S'03-M'14)}} was born in Athens, Greece, in 1986. He received the Diploma Degree in in Electrical and Computer Engineering from the National Technical University of Athens (NTUA) in 2010, and the Ph.D. degree in electrical engineering from the Interdisciplinary center for Security and Trust, (SnT), University of Luxembourg, in 2014. Dr. Christopoulos' research interests include  signal processing for satellite communications and  optimization methods for multiuser MIMO communications.
\end{IEEEbiography}

\begin{IEEEbiography}[{\includegraphics[width=1.0in,height=1.25in,clip,keepaspectratio]{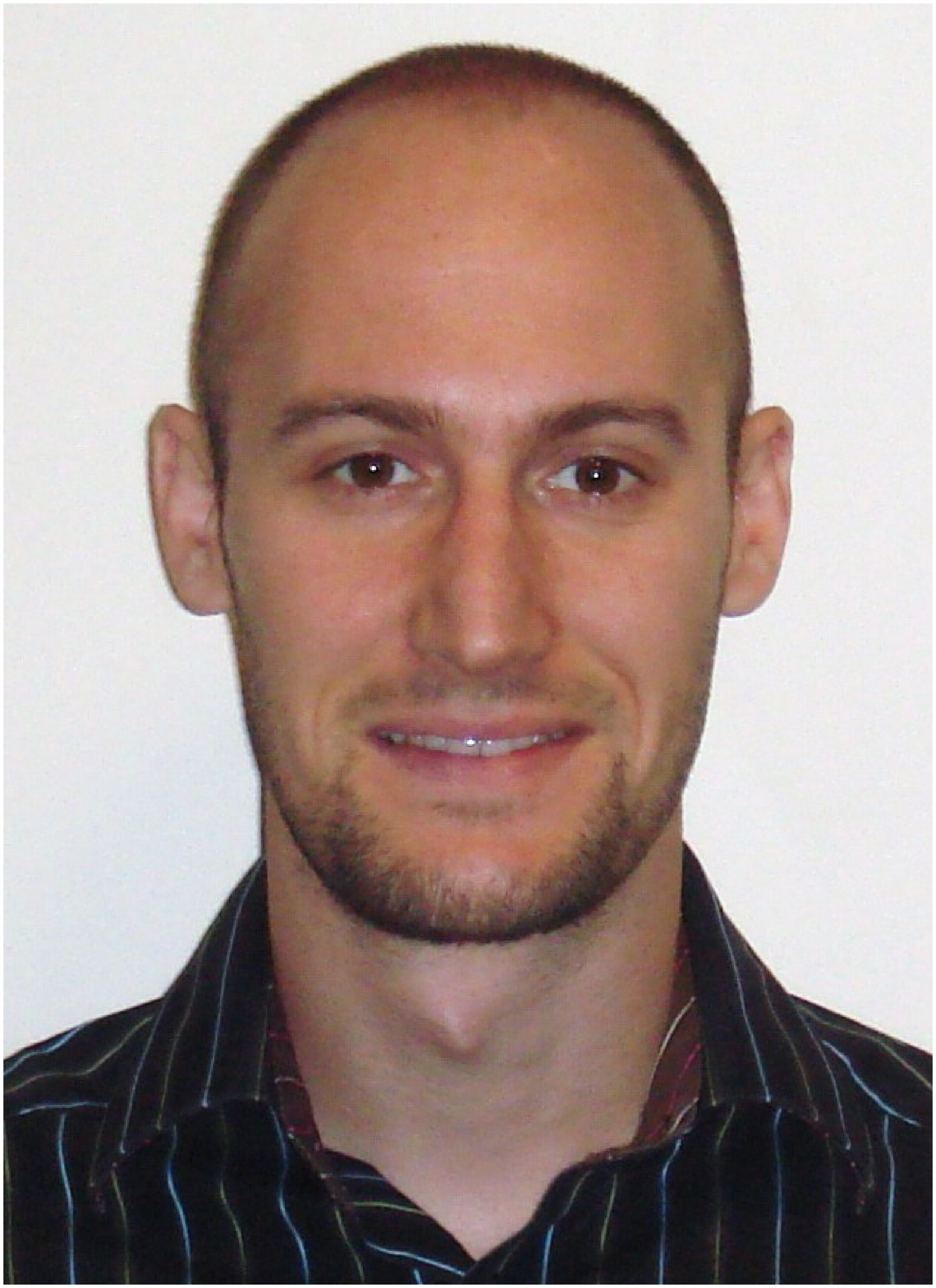}}]{\textbf{Symeon Chatzinotas (S'06-M'09-SM'13)}} received the M.Eng. degree in telecommunications from Aristotle University of Thessaloniki, Thessaloniki, Greece, in 2003 and the M.Sc. and the Ph.D. degrees in electronic engineering from the University of Surrey, Surrey, U.K., in 2009. He is currently a Research Scientist with the Interdisciplinary Centre for Security, Reliability and Trust, University of Luxembourg.

In the past, he has worked on numerous research and development projects for the Institute of Informatics and Telecommunications, National Center for Scientific Research Demokritos, Athens, Greece; the Institute of Telematics and Informatics, Center of Research and Technology Hellas, Thessaloniki, Greece; and the Mobile Communications Research Group, Center of Communication Systems Research, University of Surrey.
He is the author of more than 110 technical papers in refereed international journals, conferences, and scientific books and he is currently coediting a book on "Cooperative and Cognitive Satellite Systems".
His research interests include multiuser information theory, cooperative and cognitive communications, and transceiver optimization for terrestrial and satellite networks.
\end{IEEEbiography}

\begin{IEEEbiography}[{\includegraphics[width=1.0in,height=1.25in,clip,keepaspectratio]{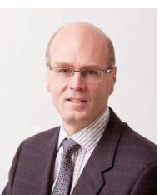}}]{\textbf{Bj$\ddot{\mathrm{o}}$rn Ottersten (S'87-M'89-SM'99-F'04)}} was
born in Stockholm, Sweden, in 1961. He received
the M.S. degree in electrical engineering and applied
physics from Link$\ddot{\mathrm{o}}$ping University, Link$\ddot{\mathrm{o}}$ping,
Sweden, in 1986 and the Ph.D. degree in electrical
engineering from Stanford University, Stanford, CA,
in 1989. \\
Dr. Ottersten has held
research positions at the Department of Electrical
Engineering, Link$\ddot{\mathrm{o}}$ping University, the Information
Systems Laboratory, Stanford University, the
Katholieke Universiteit Leuven, Leuven, and the
University of Luxembourg. During 96/97, he was Director of Research
at ArrayComm Inc, a start-up in San Jose, California based on Ottersten'
s patented technology. He has co-authored journal papers that received the
IEEE Signal Processing Society Best Paper Award in 1993, 2001, 2006, and
2013 and 3 IEEE conference papers receiving Best Paper Awards. In 1991, he
was appointed Professor of Signal Processing at the Royal Institute of Technology
(KTH), Stockholm. From 1992 to 2004, he was head of the department
for Signals, Sensors, and Systems at KTH and from 2004 to 2008, he was dean
of the School of Electrical Engineering at KTH. Currently, he is
Director for the Interdisciplinary Centre for Security, Reliability and Trust at
the University of Luxembourg. As Digital Champion of Luxembourg, he acts
as an adviser to European Commissioner Neelie Kroes. \\
Dr. Ottersten has served
as Associate Editor for the IEEE TRANSACTIONS ON SIGNAL PROCESSING and
on the editorial board of \textit{IEEE Signal Processing Magazine}. He is currently
editor in chief of \textit{EURASIP Signal Processing Journal} and a member of
the editorial boards of \textit{EURASIP Journal of Applied Signal Processing} and
\textit{Foundations and Trends in Signal Processing}. He is a Fellow of the
IEEE and EURASIP and a member of the IEEE Signal Processing Society
Board of Governors. In 2011, he received the IEEE Signal Processing Society
Technical Achievement Award. He is a first recipient of the European Research
Council advanced research grant. His research interests include security and trust, reliable wireless communications, and statistical signal processing.
\end{IEEEbiography}
\end{document}